\RequirePackage{fix-cm}
\documentclass[smallcondensed]{svjour3}     % onecolumn (ditto)
\smartqed  % flush right qed marks, e.g. at end of proof
\usepackage{graphicx}
\usepackage{natbib}
\usepackage{amsmath, amssymb}
\usepackage{algorithm}
\usepackage{algorithmic}
\usepackage{tikz}
\usepackage[linkcolor=blue,citecolor=blue,colorlinks]{hyperref}
\newcommand{\ubar}[1]{\text{\b{$#1$}}}
\newcommand{\vv}[1]{\stackrel{\text{\protect\raisebox{-4pt}[1pt][0pt]{$\mathchar"017E$}}}{#1}}
\newcommand{\vvj}{\text{\protect\raisebox{-1pt}{$\stackrel{{\text{\protect\raisebox{-4pt}[1pt][0pt]{$\mathchar"017E$}}}}{\jmath}$}}}
\newcommand{\e}{\mathrm{e}}
\newcommand{\ii}{\mathrm{i}}

\newcommand{\hvec}[1]{\hat{\vec {#1}}}
\newcommand{\mycomment}[1]{{\em // #1 }}
\DeclareMathOperator{\MIN}{MIN}
\DeclareMathOperator{\MAX}{MAX}
\newcommand{\mytiny}[1]{\fontsize{6}{7}\selectfont #1}

\begin{document}

\title{Complete spin and orbital evolution of close-in bodies using a
Maxwell viscoelastic rheology }

\author{Gwena\"el Bou\'e         \and
        Alexandre C. M. Correia  \and
        Jacques Laskar
}

\authorrunning{G. Bou\'e et al.} % if too long for running head

\institute{G. Bou\'e  \at
              \email{gwenael.boue@obspm.fr}
           \and
           G. Bou\'e \and J. Laskar \at
              IMCCE, Observatoire de Paris, UPMC Univ. Paris 6, Paris, France
           \and
           A.C.M. Correia \at
              CIDMA, Departamento de F\'isica, Universidade de Aveiro,
Campus de Santiago, 3810-193 Aveiro, Portugal
}

\date{Received: date / Accepted: date}

\graphicspath{ {figures/} }

\maketitle

\begin{abstract}
In this paper, we present a formalism designed to model tidal
interaction with a viscoelastic body made of Maxwell material.  Our
approach remains regular for any spin rate and orientation, and for any
orbital configuration including high eccentricities and close
encounters. The method is to integrate simultaneously the rotation and
the position of the planet as well as its deformation. We provide the
equations of motion both in the body frame and in the inertial frame.
With this study, we generalize preexisting models to the spatial case
and to arbitrary multipole orders using a formalism taken from quantum
theory.  We also provide the vectorial expression of the secular tidal
torque expanded in Fourier series.  Applying this model to close-in
exoplanets, we observe that if the relaxation time is longer than the
revolution period, the phase space of the system is characterized by the
presence of several spin-orbit resonances, even in the circular case. As
the system evolves, the planet spin can visit different spin-orbit
configurations.  The obliquity is decreasing along most of these
resonances, but we observe a case where the planet tilt is instead
growing. These conclusions derived from the secular torque are
successfully tested with numerical integrations of the instantaneous
equations of motion on HD~80606\,b. Our formalism is also well adapted
to close-in super-Earths in multiplanet systems which are known to have
non-zero mutual inclinations.
\keywords{Restricted Problems \and Extended Body \and Dissipative Forces
\and Planetary Systems \and Rotation}
\end{abstract}

%\tableofcontents

\section{Introduction}
\label{intro}
Short period exoplanets are tidally distorted by their stars. This
phenomenon alter both the planet rotation and its orbital evolution over
long timescale. The mechanism is the same as in the problem of a satellite 
orbiting a planet which has been modeled by \citet{Darwin_PTRSL_1880}
and generalized by \citet{Kaula_RvGSP_1964}. 

In these models, the gravitational potential of the deformed planet is
expanded in multipoles and then expressed in terms of elliptical
elements as a Fourier series truncated in eccentricity. Each term
involves a Love number associated to the amplitude of the tide and a
phase lag accounting for the non-instantaneous deformation of the
planet. These lags have been interpreted as {\em constant geometric lag}
angles \citep{MacDonald_RvGSP_1964}. However, the tidal torque should
vanish at equilibrium, i.e. when the perturbing body (star or satellite)
has a circular orbit in the planet equatorial plane with a mean motion
equal to the planet rotation speed. To remedy this problem,
\citet{Singer_GJRAS_1968} proposed a frequency-dependent theory of tides
which is now known as the {\em constant time lag model}.
According to this theory, the deformation of the planet at time $t$ is
aligned with the position occupied by the disturbing body at time
$t-\Delta t$ in the planet reference frame.

The constant time lag model has been widely used because of its intuitive
physical interpretation and also because the analytical expressions of
the tidal force and torque expanded in first order in $\Delta t$ are very 
compact and not truncated in eccentricity \citep{Mignard_MP_1979}.

More generally, Love numbers and phase lags depend on the structure and
the rheology of the planet \citep[e.g.,][]{Efroimsky_CeMDA_2012}, but
none of the two models quoted above corresponds to a physical rheology
\citep{Efroimsky_Makarov_ApJ_2013}. The constant time lag model can
nevertheless be seen as a first order expansion of a viscoelastic
rheology (\citealp[p.~740 \S{} 7]{Darwin_PTRSL_1880}; see also
\citealp{Ferraz-Mello_CeMDA_2013}).

Different rheologies have been suggested for rocky and giant gaseous
planets \citep[e.g.,][]{Ogilvie_Lin_ApJ_2004,Efroimsky_Lainey_JGRE_2007,
Henning_etal_ApJ_2009,Remus_etal_AA_2012, Efroimsky_ApJ_2012}. A few of
them have been proposed because of their (mathematical and physical)
simplicity, others are motivated by laboratory and/or numerical
experiments or by geophysical measurements. 

In the general case, mathematical models describing the rheology are
intricate and do not allow to follow the long term rotation and orbital
motion without a Fourier series as in \citeauthor{Kaula_RvGSP_1964}'s
theory. This is a disadvantage since such expansions are only valid at low
eccentricities unless a huge number of terms is kept (see the discussion
in the Appendix of \citet{Ferraz-Mello_CeMDA_2013}).

A few physical rheologies can nevertheless be treated without Fourier
series, such as the viscous creep model \citep{Ferraz-Mello_CeMDA_2013}
and the Maxwell viscoelastic model \citep{Correia_AA_2014}. It should
be noted that dissipation is equivalent in both models
\citep{Correia_AA_2014, Ferraz-Mello_CeMDA_2015}. These rheologies can
be seen as first order low-pass filters and can thus be modeled by first
order differential equations. In these models, coefficients of the
potential are integrated at the same time as the orbital and rotational
elements.  There is no requirement regarding the perturbation : it does
not have to be periodic nor low-eccentric.

Recently, \citet{Frouard-arXiv-2016} proposed an alternative approach
with the same advantages where the extended body is made of a large
number $N$ of massive gravitating particles linked by damped massless
springs. The demo version of this method, described in Ibid., employed
springs obeying the Kelvin-Voigt law. Accordingly, the resulting shear
response of the mesh was close to Kelvin-Voigt. By choosing different
deformation laws for the springs, it is possible to endow the mesh with
different rheologies. This approach can easily be set up to model bodies
with complex internal structure and/or geometry. But it requires the
integration of about $6N$ differential equations. 

Earths and super-Earths are assumed to behave like a Maxwell body at low
frequency, but in the opposite regime such model does not account for
enough dissipation and an Andrade rheology is required
\citep{Efroimsky_ApJ_2012}. This composite model, which can only be
expressed mathematically as a truncated Fourier series, led to unexpected
results. Indeed, according to \citeauthor{Singer_GJRAS_1968}'s
and \citeauthor{Mignard_MP_1979}'s constant $\Delta t$ model, the
rotation of a planet without permanent quadrupole on eccentric orbit is
expected to be pseudo(or super)synchronous, while with this new rheology
the only stable configurations are at the vicinity of spin-orbit
resonances \citep{Makarov_Efroimsky_ApJ_2013}. 
Actually, entrapment into spin-orbit resonances is not an exclusive property
of the composite Maxwell + Andrade rheology but a robust entailment of
linear rheologies \citep{Makarov_Efroimsky_ApJ_2013}. In particular,
these resonances are also expected in the case of purely Maxwell bodies
\citep{Correia_AA_2014}.

In summary, Maxwell rheology presents two advantages : a simple
mathematical representation valid at all eccentricities and similar
qualitative outcomes as more complex models.
In \citep{Ferraz-Mello_CeMDA_2013, Correia_AA_2014}, the problem
has been studied in the planar case where the spin of the planet is
orthogonal to the orbital plane. In this work we present a formalism for
inclined systems. For that purpose, multipole expansion in complex
spherical harmonics $Y_{l,m}$, as initiated by
\citet{Mignard_CeMec_1978}, proves to be efficient especially as these
functions have simple expressions in terms of Cartesian coordinates,
they are the eigenvectors of ladder operators from which the tidal force
and torque are derived, and each operation (rotation, differentiation,
...) on these functions can be found in any textbook about quantum
mechanics such as in \citep{Varshalovich_book_1988}.  The equations of
motion are given both in the frame of the planet, in which tides are
naturally expressed, but also in a fixed reference frame more suitable
for describing the orbital evolution. In this work, we mainly
concentrate on the instantaneous equations of motion valid at all
eccentricities, except in Section~\ref{sec.secular} where we provide the
secular tidal torque in the form of a standard Fourier expansion.

The paper is organized as follows : the model and the notations are
presented in Section~\ref{sec:model}; the next two sections
(\ref{sec:planetframe} and \ref{sec.F0}) provide the instantaneous
equations of motion in the body frame and in the inertial frame;
Section~\ref{sec.secular} focuses on the secular evolution. It provides
the secular torque and maps of the secular evolution of the spin-axis; our
model is then applied on HD~80606\,b in Section~\ref{sec:results}; the
conclusion is drawn in Section~\ref{sec:conclusion}.

\section{Model and notation}
\label{sec:model}
We wish to determine the orbital and rotational evolution of an extended
planet of mass $m_1$ orbiting a point-mass star $m_0$. The planet
is assumed to be made of a viscoelastic fluid governed by Maxwell
rheology. At rest, the planet would thus be a perfect sphere of radius
$R$. In this problem, the planet is deformed by its rotation around its
spin-axis and by the differential gravitational field of the star. 

We denote by $V(\vv x, t)$ the gravitational potential of the deformed
planet at time $t$ and at the position $\vv x$ with respect to its
center of mass. In the following, we provide the expression of this
potential and the equations of motion both in the body frame ${\cal
F}_p$ rotating with the planet and in an inertial frame ${\cal F}_0$. 

Thus, for any vector $\vv x$ in the physical space written with
an arrow, we distinguish its coordinates in ${\cal F}_0$ represented by
a bold lower case such as $\vec x$ from those in ${\cal F}_p$ denoted by
a bold face capital letter such as $\vec X$. We also let $x=X=\|\vv x\|$
be its norm. Unit vectors are denoted with a hat, e.g., $\hat x = \vv
x/x$.

Let $f(\vv x, t)$ be an arbitrary function whose expressions in the
frames ${\cal F}_0$ and ${\cal F}_p$ are respectively denoted by
$f_0(\vec x, t)$ and $f_p(\vec X, t)$. We define the gradient operators
$\vec \nabla_{\vec x}$ and $\vec \nabla_{\vec X}$ by
$$
\vec \nabla_{\vec x} f(\vv x, t) \equiv \vec \nabla f_0(\vec x, t)
\qquad\text{and}\qquad
\vec \nabla_{\vec X} f(\vv x, t) \equiv \vec \nabla f_p(\vec X, t)\ .
$$
Equivalently, we consider the angular momentum operators $\vec J_{\vec
x}$ and $\vec J_{\vec X}$ in the frames ${\cal F}_0$ and ${\cal F}_p$,
respectively, such that
$$
\vec J_{\vec x} f(\vv x, t) = \vec J f_0(\vec x, t) \equiv 
-\ii\,\vec x \times \vec \nabla f_0(\vec x, t)
$$
and
$$
\vec J_{\vec X} f(\vv x, t) = \vec J f_p(\vec X, t) \equiv 
-\ii\, \vec X \times \vec \nabla f_p(\vec X, t)
$$
where $\ii = \sqrt{-1}$. The gradient and the angular momentum operators
will be used to express the tidal force and torque, respectively.

The formalism described in this paper is completely vectorial and can
thus be computed in any coordinate system (either spherical or Cartesian).
We have chosen the complex Cartesian coordinate system as defined in
\citep{Varshalovich_book_1988} because it leads to very compact
formulas. This system is defined as follows, for any vector $\vv v$,
its coordinates in ${\cal F}_0$ are $\vec v = (v_+, v_0, v_-)$ with
$$
v_+ = - \frac{1}{\sqrt 2} (v_x + \ii v_y)\ ,
\qquad
v_0 = v_z\ ,
\qquad
v_- = \frac{1}{\sqrt 2} (v_x - \ii v_y)\ ,
$$
where $(v_x, v_y, v_z)$ are the usual real Cartesian coordinates. The
coordinates $\vec V = (V_+, V_0, V_-)$ in ${\cal F}_p$ are equivalently 
defined using the same rule. For any complex quantity $z\in\mathbb{C}$,
the complex conjugate is written with a bar as $\bar z$. We stress that
$v_- = -\bar v_+$ and thus a vector $\vv v$ is fully characterized by
only two components, e.g., $v_+$ and $v_0$.

\section{Description in the planet frame}
\label{sec:planetframe}

\subsection{Tidal potential}
The gravitational potential $V(\vv x, t)$ of the planet is the sum of
two components: the potential at rest $V^0(\vv x) = -G m_1 / x$ and a
small correction $V'(\vv x, t)$ due to the mass redistribution
within the planet. The latter is usually expressed in the body frame
${\cal F}_p$. Outside of the planet, i.e. for $\|\vec x\|>R$, $V'$ satisfies Laplace's equation $\Delta V' = 0$
and remains finite when $\|\vv x\|\rightarrow\infty$. Thus, it can be expanded
in spherical harmonics $Y_{l, m}$ (here we use the Schmidt
semi-normalization convention, see Appendix~\ref{sec.sh}).
Beyond the planet surface, we have then
$$
V'(\vv x, t) = \sum_{l=2}^\infty V'_l(\vv x, t)
$$
with
\begin{equation}
V'_l(\vv x, t) = -\frac{Gm_1}{R} \left(\frac{R}{X}\right)^{l+1} \sum_{m=-l}^l
 \bar Z_{l,m}(t) Y_{l,m}(\hvec X)
\label{eq.V'ell}
\end{equation}
where $Z_{l,m}(t)$ are coefficients whose relation to {\em Stokes
coefficients} will be detailed later on. This deformation is induced by
a ``disturbing potential'' $W(\vv x, t)$ associated to the rotation of
the planet and to the differential potential of the star. Let $\vv
\omega$ be the instantaneous rotation vector of the planet, and $\vec
\Omega$ and $\vec \omega$ its coordinates in ${\cal F}_p$ and ${\cal
F}_0$, respectively. If we neglect the radial term of the centrifugal
force which has no effect if the planet is made of incompressible fluid,
both disturbing potentials can also be expanded in spherical harmonics
$W(\vv x, t) = \sum_{l=2}^\infty W_l(\vv x, t)$, with
\begin{subequations}
\begin{align}
W_2(\vv x, t) =& \frac{1}{3} \Omega^2(t) X^2 \sum_{m=-2}^2 \bar
Y_{2,m}(\hvec \Omega(t)) Y_{2, m}(\hvec X) \nonumber \\
& - G m_0 
  \frac{X^2}{X_\star^3(t)} \sum_{m=-2}^2
\bar Y_{2,m}(\hvec X_\star(t)) Y_{2,m}(\hvec X)
\end{align}
and for $l\geq3$,
\begin{equation}
W_l(\vv x, t) = -G m_0
  \frac{X^l}{X_\star^{l+1}(t)} \sum_{m=-l}^l
\bar Y_{l,m}(\hvec X_\star(t)) Y_{l,m}(\hvec X)
\end{equation}%
\label{eq.Well}%
\end{subequations}
where $\vec X_\star(t)$ is the coordinates in the frame ${\cal F}_p$ of
the position $\vv x_\star(t)$ of the star relative to the planet
barycenter at time $t$.  To simplify the notation, the explicit time
dependency of $\vec \Omega(t)$ and $\vec X_\star(t)$ will be dropped in
the following equations.

According to the linear model of tides, at $\|\vv x_R\| = R$ from the
planet center, 
$V'_l(\vv x_R, t)$ is a linear
combination of all
$W_l(\vv x_R, t')$ with $t' \leq t$. Thus, for all $l\geq 2$,
\begin{equation}
V'_l(\vv x_R, t) = k_l(t) * W_l(\vv x_R, t)
= \int_{-\infty}^t k_l(t-t') W_l (\vv x_R, t')\,
dt'
\ ,
\label{eq.convol}
\end{equation}
where $k_l(t)$ is a {\em Love distribution} such that $k_l(t) = 0$
for all $t>0$ and where $*$ is the convolution product. The terminology
is chosen by analogy with the Love numbers $k_l$. Note that in
\citep{Efroimsky_CeMDA_2012}, these distributions are noted $\overset{\tikz \fill 
(0,0) rectangle (0.035,0.035);}{k}_l(t)$. Love distributions are a
property of the planet. They depend on its internal structure and
composition, but not on the perturbing body. Substituting in
(\ref{eq.convol}) the expressions (\ref{eq.V'ell}) and (\ref{eq.Well})
of $V'_l$ and $W_l$ respectively, we get
\begin{equation}
Z_{l,m}(t) = k_l(t) * Z^\star_{l,m}(t)\ ,
\label{eq.convolZ}
\end{equation}
with
\begin{subequations}
\begin{equation}
Z^\star_{2,m}(t) = -\frac{1}{3} \frac{\Omega^2 R^3}{G m_1} Y_{2,m}(\hvec
\Omega) + \frac{m_0}{m_1} \left(\frac{R}{X_\star}\right)^{3}
Y_{2,m}(\hvec X_\star)
\end{equation}
and for all $l \geq 3$,
\begin{equation}
Z^\star_{l,m}(t) = \frac{m_0}{m_1} \left(\frac{R}{X_\star}\right)^{l+1}
Y_{l,m}(\hvec X_\star)\ .
\end{equation}%
\label{eq.zstar}%
\end{subequations}

\subsection{Differential equations satisfied by the $Z_{l,m}$}
The convolution equations (\ref{eq.convol}) and
(\ref{eq.convolZ}) are very general. They only assume that the tidal
response is linear and isotropic in the frame of the planet. Now, we add
a new hypothesis in the model saying that the planet behaves like an
homogeneous viscoelastic body with Maxwell rheology. In that case, the
Fourier transform $\ubar k_l$ of the distribution $k_l$ is of
the form\footnote{Note that if the material composing the extended body
was governed by the Newtonian creep rheology or by the Kelvin-Voigt one, 
$\ubar k_l(\nu)$ would have the same analytical expression but with 
$\tau_e = 0$.} \citep[e.g.,][]{Henning_etal_ApJ_2009}
\begin{equation}
\ubar k_l(\nu) = k_l^0 \frac{1+\ii\tau_e\nu}{1+\ii\tau_l \nu}
\label{eq.Fourier}
\end{equation}
where $k_l^0 = 3/[2(l-1)]$ is the fluid Love number of
degree $l$, $\tau_l = (1+A_l)\tau_e$ is a global relaxation
time, $\tau_e = \eta/\mu$ is the elastic or Maxwell relaxation time,
$A_l \tau_e = (2l^2+4l+3)\eta/(l g \rho R)$ is the fluid
relaxation time, $\eta$ is the viscosity, $\mu$ is the rigidity (or
shear modulus), and $\rho$ is the mean density. It must be stressed that
the aforementioned expressions of $k^0_l$ and $\tau_l$ only hold
for perfectly homogeneous incompressible viscous sphere. Real planets
are stratified and thus each $k^0_l$, $\tau_l$, and even $\tau_e$
can be considered as free parameters that have to be fitted to reproduce
the response of a more complex internal structure
\citep[e.g.,][]{Peltier_RvGSP_1974}.

Given the expression of the Fourier transform of $k_l$
(Eq.~\ref{eq.Fourier}), the convolution equation (Eq.~\ref{eq.convolZ})
becomes a first order differential equation 
\citep{Correia_AA_2014}
$$
Z_{l,m} + \tau_l \dot Z_{l,m} = Z^e_{l,m} + \tau_e \dot
Z^e_{l,m}
$$
where $Z^e_{l,m} = k_l^0 Z^\star_{l,m}$. Following
\citet{Ferraz-Mello_CeMDA_2015}, we can also express the previous equation
in a simpler form that does not depend on the derivatives of
$Z^e_{l,m}$ as
\begin{equation}
Z_{l,m} = \left(1-\frac{\tau_e}{\tau_l}\right) Z^\nu_{l,m} +
\frac{\tau_e}{\tau_l}Z^e_{l,m}
\qquad \mathrm{with}\qquad
Z^\nu_{l,m} + \tau_l \dot Z^\nu_{l,m} = Z^e_{l,m}
\ .
\label{eq.zze}
\end{equation}
We recall that all $Z^e_{l,m}(t)=k^0_l Z^\star_{l,m}(t)$, given
by Eq.~(\ref{eq.zstar}), are only functions of the instantaneous
rotation vector $\vec \Omega(t)$ of the planet and of the position $\vec
X_\star(t)$ of the disturbing star at time $t$. There is no restriction
regarding the orbital evolution. Equation~(\ref{eq.zze}) can thus be
integrated even if the trajectory is chaotic, aperiodic, or highly
eccentric.

\subsection{Stokes coefficients and matrix of inertia}

Conventionally, the potential is developed in the body frame as
\citep[e.g.,][]{Lambeck_book_1988}
$$
V(\vv x, t) = -\frac{Gm_1}{R} \sum_{l=0}^\infty
\left(\frac{R}{X}\right)^{l+1} 
 \sum_{m=0}^l
(C_{l,m} \cos(m\phi) + S_{l,m}
\sin(m\phi) ) P_{l,m}(\cos \theta)\ ,
$$
where $C_{l,m} = C^0_{l,m} + C'_{l,m}$ and $S_{l,m} = S^0_{l,m} +
S'_{l,m}$ are the Stokes coefficients splitted into their permanent part
(superscript 0) and their deformation part (with a prime), and where
$(\phi, \theta)$ are the longitude and colatitude of $\vv x$ in ${\cal
F}_p$. In our problem, $C^0_{l,m}=S^0_{l,m}=0$ because the body
is assumed to be spherical without tidal or rotational deformation. We
thus have $C_{l,m}=C'_{l,m}$ and $S_{l,m}=S'_{l,m}$. A comparison with
the equation (\ref{eq.V'ell}) using the definition of the spherical
harmonics given in Appendix~\ref{sec.sh} shows that
\begin{equation}
Z_{l,m} =
 (-1)^m \frac{1+\delta_{m,0}}{2} \sqrt{\frac{(l+m)!}{(l-m)!}}\big(C_{l,m} + \ii S_{l,m}\big)
\qquad \mathrm{if}\quad m\geq 0\ .
\label{eq.zlm}
\end{equation}
In this expression, $\delta_{i,j}$ is Kronecker's delta equal to 1 if $i=j$
and 0 otherwise.
The other coefficients are given by $Z_{l,m} = (-1)^m \bar
Z_{l,-m}$\ .

The relation (\ref{eq.zlm}) between the coefficients $Z_{l,m}$ and
Stokes coefficients allows to compute the matrix of inertia $\tens
I_n(Z_{l,m})$. To express the result, let us first denote by $\xi$ the
normalized moment of inertia such that, without deformation, $\tens I_n
= \xi m_1 R^2 \tens I_{3\times3}$ where $\tens I_{3\times3}$ is the
identity. For homogeneous body, we have $\xi = 2/5$, but more generally,
$\xi$ is related to the fluid Love number $k_2^0$ through the
Darwin-Radau equation \citep[e.g.,][]{Jeffreys_1976}
$$
\xi = \frac{2}{3}\left(1-\frac{2}{5}\sqrt{\frac{4-k_2^0}{1+k_2^0}}\right)\ .
$$ 
Once the planet is deformed by its rotation and by tides, we have to add
in the matrix of inertia a contribution due to the mass redistribution
within the planet, and we get
$$
\tens I_n(Z_{l,m}) = 
\left(\xi m_1R^2\right)\tens I_{3\times3} + m_1R^2 \begin{pmatrix}
\displaystyle  \frac{1}{3}        Z_{2, 0}       & 
\displaystyle -\frac{1}{\sqrt 3}  Z_{2, 1}       & 
\displaystyle  \sqrt{\frac{2}{3}} Z_{2, 2}       \\[1.2em]
\displaystyle  \frac{1}{\sqrt 3}  Z_{2,-1}       & 
\displaystyle -\frac{2}{3}        Z_{2, 0}       & 
\displaystyle  \frac{1}{\sqrt 3}  Z_{2, 1}       \\[1.2em]
\displaystyle  \sqrt{\frac{2}{3}} Z_{2,-2}       &
\displaystyle -\frac{1}{\sqrt 3}  Z_{2,-1}       &
\displaystyle  \frac{1}{3}        Z_{2, 0}
\end{pmatrix}\ .
$$
This matrix of inertia is
complex because it is defined such that the angular momentum $\vec L$
reads
$$
\begin{pmatrix}
L_+ \\ L_0 \\ L_-
\end{pmatrix} = \tens I_n(Z_{l,m}) \begin{pmatrix}
\Omega_+ \\ \Omega_0 \\ \Omega_-
\end{pmatrix}\ .
$$
The modification of the matrix of inertia due the mass redistribution is
a small correction. In the subsequent simulations, the rotation vector
$\vv \omega$ is deduced from the angular momentum $\vv \ell$ through the
relation $\vv \ell = C \vv \omega$ with $C=\xi m_1R^2$ as in
\citep{Correia_AA_2014}.

\subsection{Complete set of differential equations}
\label{sec.eomZ}
Given the gravitational potential $V(\vv x, t)$ raised by the planet,
the force acting on the star is $\vec F = -m_0 \vec \nabla_{\vec X}
V(\vv x, t)$. If the reference frame were not rotating, we
would have formally obtained the orbital evolution of the system
with $\ddot{\vec X} = \vec F/\beta$, where $\beta = m_0m_1/(m_0+m_1)$ is
the reduced mass.  Here, we have to add the usual inertial forces. We
get
$$
\ddot{\vec X} = -\frac{m_0}{\beta} \vec \nabla_{\vec X} V(\vv x, t)
- \dot{\vec \Omega} \times \vec X - 2 \vec \Omega \times \dot{\vec X}
-\vec \Omega \times (\vec \Omega \times \vec X)\ .
$$
In order to have first order differential equations, we introduce the
velocity $\vec U = \dot{\vec X} + \vec \Omega\times\vec X$ of the star
relative to the planet center of mass in the frame ${\cal F}_0$. We have
then
$$
\dot{\vec X} = \vec U - \vec \Omega \times \vec X
\qquad \text{and}\qquad
\dot{\vec U} = -\frac{m_0}{\beta} \vec \nabla_{\vec X} V(\vv x, t)
- \vec \Omega \times \vec U\ .
$$
The torque on the planet is $\vec T = -\vec X \times \vec F$. Thus, the
evolution of the angular momentum $\vec L$ of the planet in ${\cal F}_p$
is given by
$$
\dot{\vec L} = m_0 \vec X \times \vec \nabla_{\vec X} V(\vv x, t) -
\vec \Omega \times \vec L\ .
$$
Now, we substitute the expression of $V(\vv x, t)$ and we add the
equation of motion satisfied by $Z^\nu_{l,m}$. The result is
\begin{subequations}
\begin{align}
&\dot{\vec X} = \vec U - \vec \Omega \times \vec X\ , \\
&\dot{\vec U} = -G(m_0+m_1) \left(\frac{\hvec X}{X^2} 
               - \sum_{l=2}^{l_\mathrm{max}} R^l \sum_{m=-l}^l
                \bar Z_{l,m} \vec \nabla\left(\frac{Y_{l,m}(\hvec
                X)}{X^{l+1}}\right)\right) - \vec \Omega \times \vec U\ , \\
&\dot{\vec L} = -\ii \frac{Gm_0m_1}{X}
\sum_{l=2}^{l_\mathrm{max}} \left(\frac{R}{X}\right)^{l}
\sum_{m=-l}^l \bar Z_{l,m} \vec J \left(Y_{l,m}(\hvec X)\right)
                - \vec \Omega \times \vec L\ , 
\label{eq.Ldot} \\
&\dot Z^\nu_{l,m} = \frac{1}{\tau_l} (Z^e_{l,m} - Z^\nu_{l,m})\ ,
\quad l \in \{2, l_\mathrm{max}\}\ , \quad m \in \{0,l\}\ ,
\label{eq.dZnu}
\end{align}%
\label{eq.evolZ}%
\end{subequations}
where $l_\mathrm{max}$ is the maximal order at which the multipole
expansion is performed. For this problem, the state vector is
$\vec Y = (X_0, X_+, U_0, U_+, L_0, L_+, Z^\nu_{l,m})$ with $2\leq
l \leq l_\mathrm{max}$ and $0\leq m \leq l$. Auxiliary
quantities are computed as follows:
\begin{itemize}
\item $X_- = -\bar X_+$, $U_- = -\bar U_+$, $L_- = -\bar L_+$,
\item $X=\|\vv x\| = X_0^2 - 2X_-X_+$,
\item $\vec \Omega = C^{-1}\vec L$,
\item $Y_{l,m}(\hvec X)$ with $l\in\{2,l_\mathrm{max}+1\}$ and $m\in\{-l,l\}$
from Appendix~\ref{sec.sh},
\item $\vec \nabla(Y_{l,m}(\hvec X)/X^{l+1})$ and $\vec
J(Y_{l,m}(\hvec X))$ with $l\in\{2,l_\mathrm{max}\}$ and $m \in \{-l,l\}$ from Appendix~\ref{sec.lo},
\item $(Z_{l,m})_{m\geq 0}$ from Eq.~(\ref{eq.zze}) and $Z_{l,-m}=(-1)^m\bar Z_{l,m}$,
\item $Z^e_{l,m}=k_l^0 Z^\star_{l,m}$ with $Z^\star_{l,m}$ given by Eq.~(\ref{eq.zstar}).
\end{itemize}
We stress that the state vector contains the minimal set of variables
allowing to integrate the problem. Indeed, $(X_0, U_0, L_0,
Z^\nu_{l,0})$ are real and the others are complex. We thus have six
(real) coordinates for the orbit: position and velocity, three for the
angular momentum but none for the orientation (because the body is
spherical at rest), and $2l+1$ coefficients per multipole of degree
$l$. However, this formalism is not the most convenient to study 
$n$-body problems because trajectories are followed in the frame of the
tidally deformed planet rather than in the inertial frame. Moreover, if 
more than one body is allowed to be distorted, one also has to
integrate orientations to compute change of bases. This increases the
dimension of the state vector. In the next section, we provide an
alternative approach directly written in the inertial frame ${\cal
F}_0$.

\section{Description in the inertial frame}
\label{sec.F0}

\subsection{Tidal potential}
\label{sec.potentialz}
In the previous section, we wrote the harmonics of the additional
potential $V'_l(\vv x, t)$ in the body frame as
$$
V'_l(\vv x, t) = -\frac{Gm_1}{R}\left(\frac{R}{X}\right)^{l+1}
\sum_{m=-l}^l \bar Z_{l,m}(t) Y_{l,m}(\hvec X)\ ,
$$
but we could also have decomposed $V'_l(\vv x, t)$ in the inertial
frame as
$$
V'_l(\vv x, t) = -\frac{Gm_1}{R}\left(\frac{R}{x}\right)^{l+1}
\sum_{m=-l}^l \bar z_{l,m}(t) Y_{l,m}(\hvec x)
$$
with new time-dependent coefficients $z_{l,m}(t)$ expressing the
gravity field of the planet in the inertial frame. Coefficients
$Z_{l,m}$ and $z_{l,m}$ are related between themselves through 
Wigner's D matrix of size $(2l+1)\times(2l+1)$ denoted 
$\tens D^l_{m,m'}(t)$ and associated to the orientation of the frame
${\cal F}_p$ with respect to ${\cal F}_0$ at time $t$.
By definition, we have
\begin{equation}
Y_{l,m}(\hvec X) = \sum_{m'=-l}^l \tens D^l_{m',m}(t)
Y_{l,m'}(\hvec x) \quad\mathrm{thus}\quad
z_{l,m}(t) = \sum_{m'=-l}^l \bar{\tens D}^l_{m,m'}(t)
Z_{l,m'}(t)\ .
\label{eq.Wigner}
\end{equation}
For the present study, we do not need to explicit this matrix. We refer
the interested reader to the chapter 4 of
\citep{Varshalovich_book_1988}. We can nevertheless deduce the equation
of evolution of $z_{l,m}(t)$ from that of $Z_{l,m}(t)$ (see
Appendix~\ref{sec.td}). We get
$$
z_{l,m}+\tau_l \left(
\dot z_{l,m}-\ii \sum_{m'} \bar{\tens J}^l_{m,m'}(\vec \omega)z_{l,m'}
\right) = z^e_{l,m}+\tau_e \left(
\dot z^e_{l,m}-\ii \sum_{m'} \bar{\tens J}^l_{m,m'}(\vec \omega)z^e_{l,m'}
\right)\ ,
$$
where $\bar{\tens J}^l_{m,m'}(\vec \omega)$ is the complex conjugate
of the matrix  $\tens J^l_{m,m'}(\vec \omega)$ expressing the inertia
felt by the $z_{l,m}$ which are given in the fixed frame ${\cal F}_0$
rather than in the frame of the planet ${\cal F}_p$. The equilibrium
$z^e_{l,m}$ in the right-hand side are, as in the previous section,
\begin{subequations}
\begin{equation}
z^e_{2,m}(t) = k_2^0\left( -\frac{1}{3} \frac{\omega^2R^3}{Gm_1} Y_{2,m}(\hvec
\omega) + \frac{m_0}{m_1}\left(\frac{R}{x_\star}\right)^3 Y_{2,m}(\hvec
x_\star) \right)\ ,
\end{equation}
and for $l\geq 3$,
\begin{equation}
z^e_{l,m}(t) = k_l^0 \frac{m_0}{m_1} \left(\frac{R}{x_\star}\right)^{l+1}
Y_{l,m}(\hvec x_\star)\ .
\end{equation}%
\label{eq.zedef}%
\end{subequations}
Applying the change of variable proposed in
\citep{Ferraz-Mello_CeMDA_2015},
\begin{equation}
z_{l,m} = \left(1-\frac{\tau_e}{\tau_l}\right)z^\nu_{l,m}
+ \frac{\tau_e}{\tau_l} z^e_{l,m}\ ,
\label{eq.znu}
\end{equation}
we obtain the simplified equations of motion
\begin{equation}
z^\nu_{l,m} + \tau_l \left(
\dot z^\nu_{l,m} - \ii \sum_{m'} \bar{\tens J}^l_{m,m'}(\vec
\omega) z^\nu_{l,m'}
\right) = z^e_{l,m}\ .
\label{eq.dz}
\end{equation}
Hence, in ${\cal F}_0$, the time derivative of the harmonic $z_{l,m}$
is not only a function of itself and $z^e_{l,m}$, it also depends on
the other coefficients of degree $l$ but of different orders $m'$. The
system of differential equations is not diagonal anymore. This is the
price to pay when we express tides in the inertial frame.

\subsection{Matrix of inertia}
In the inertial frame, the matrix of inertia $\tens I_n(z_{l,m})$ such that $\vec
\ell = \tens I_n(z_{l,m}) \vec \omega$ has exactly the same form as in the planet
frame except that capital $Z_{l,m}$'s have to be replace by their lower
case counterparts $z_{l,m}$. The result is
$$
\tens I_n(z_{l,m}) = \left(\xi m_1R^2\right)\tens I_{3\times3} + m_1R^2 \begin{pmatrix}
\displaystyle  \frac{1}{3}        z_{2, 0}       & 
\displaystyle -\frac{1}{\sqrt 3}  z_{2, 1}       & 
\displaystyle  \sqrt{\frac{2}{3}} z_{2, 2}       \\[1.2em]
\displaystyle  \frac{1}{\sqrt 3}  z_{2,-1}       & 
\displaystyle -\frac{2}{3}        z_{2, 0}       & 
\displaystyle  \frac{1}{\sqrt 3}  z_{2, 1}       \\[1.2em]
\displaystyle  \sqrt{\frac{2}{3}} z_{2,-2}       &
\displaystyle -\frac{1}{\sqrt 3}  z_{2,-1}       &
\displaystyle  \frac{1}{3}        z_{2, 0}
\end{pmatrix}\ .
$$

\subsection{Equations of motion}
In the inertial frame, orbital and rotational equations of motion are
simply written without terms of inertia. The evolution of the gravity
field coefficients are taken from Sect.~\ref{sec.potentialz}. We get
\begin{subequations}
\begin{align}
\dot{\vec x} &= \vec u\ , \\
\dot{\vec u} &= -G(m_0+m_1) \left(\frac{\hvec x}{x^2} - \sum_{l=2}^{l_\mathrm{max}}
R^l \sum_{m=-l}^l \bar z_{l,m} 
\vec \nabla\left(\frac{Y_{l,m}(\hvec x)}{x^{l+1}}\right)\right)\ , \\
\dot{\vec \ell} &= -\ii \frac{G m_0m_1}{x} 
\sum_{l=2}^{l_\mathrm{max}} \left(\frac{R}{x}\right)^l\sum_{m=-l}^l \bar z_{l,m}
\vec J \left( Y_{l,m}(\hvec x)\right)\ , 
\label{eq.ldot}
\\
\dot z^\nu_{l,m} &= 
                    \frac{1}{\tau_l}\left(z^e_{l,m} - z^\nu_{l,m} \right)
                    +\ii \sum_{m'=-l}^l \left(\bar{\tens J}^l_{m,m'}(\vec \omega) z^\nu_{l,m'}\right)
\ , \  l \in \{2,l_\mathrm{max}\}\ , \  m \in \{0,l\}\ .
\label{eq.dznu}
\end{align}
\label{eq.evolz}
\end{subequations}
The state vector $\vec y = (x_0, x_+, u_0, u_+, \ell_0, \ell_+, 
z^\nu_{l,m})$ with $l\in\{2,l_\mathrm{max}\}$ and $m\in\{0,
l\}$ has the same dimension as in the body frame
(Sect.~\ref{sec.eomZ}). Auxiliary quantities are computed in the same
way:
\begin{itemize}
\item $x_- = -\bar x_+$, $u_-=-\bar u_+$, $\ell_- = -\bar \ell_+$, 
\item $x = \|\vec x \| = x_0^2-2x_-x_+$, 
\item $\vec \omega = C^{-1} \vec \ell$,
\item $Y_{l,m}(\hvec x)$ with $l \in \{2,l_\mathrm{max}+1\}$ and
$m\in\{-l,l\}$ from Appendix~\ref{sec.sh},
\item $\vec \nabla (Y_{l,m}(\hvec x) / x^{l+1})$ and $\vec
J(Y_{l,m}(\hvec x))$ with $l\in\{2,l_\mathrm{max}\}$ and $m\in\{-l,l\}$ 
from Appendix~\ref{sec.lo},
\item $(z_{l,m})_{m\geq 0}$ from Eq.~(\ref{eq.znu}) and
$z_{l,-m} = (-1)^m \bar z_{l,m}$,
\item $z^e_{l,m}$ from Eq.~(\ref{eq.zedef}),
\item $\tens J_{m,m'}^l(\vec \omega)$ from
Appendix~\ref{sec.td}.
\end{itemize}
This formalism has the advantage that it can easily be extended to
$n$-body problems with additional distorted planets. There is no need to
add the orientation of the extended bodies in the state vector nor to
perform change of bases. Evidently, this is not true if planets have
permanent multipoles.

\section{Secular rotation}
\label{sec.secular}
In the previous section, we have presented a set of differential
equations describing the evolution of the planet rotation, orbital
motion, and instantaneous deformation under tidal dissipation.
Nevertheless, the influence of tides on the orbit and on the planet spin
are only significant over long timescales. In this section we
propose to express the {\em secular} torque averaged over one orbital
period. Our goal is to look for the existence of any rotation
equilibria at non-zero obliquity. This torque is computed in the
inertial frame ${\cal F}_0$.

To do so, it should first be noted that the equations of motion of the
gravity field coefficients \ref{eq.dznu}) are
those of driven harmonic oscillators. The general solution is a sum of a
transient solution, which is damped within a timescale $\tau_l$, and a
steady-state proportional to the driving force. We retain the forced
solution, substitute it in the expression of the instantaneous torque
\ref{eq.ldot}), and average the result to get the secular torque
\citep[see Sects.~3,4 of][]{Correia_AA_2014}. The result is
given in the form of a Fourier series. As notified earlier, such
expansions are not suited to numerical simulations of highly eccentric
systems. The secular torque is provided here as a guideline to probe the
phase-space of the rotation motion of a single planet system on a
Keplerian orbit.

\begin{figure}
\begin{center}
\includegraphics[width=0.8\linewidth]{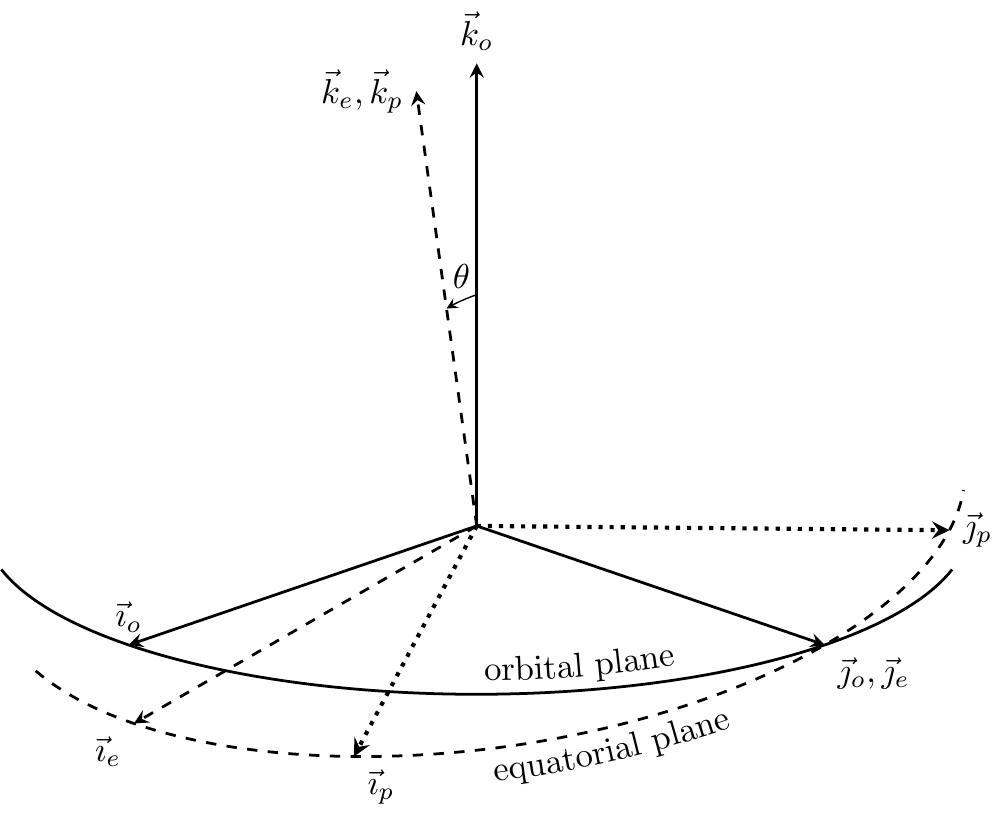}
\caption{\label{fig.frame}Definition of the basis vectors associated to the
orbit, to the equatorial plane, and to the planet frame:
the orbital basis ${\cal B}_o = (\vv \imath_o, \vvj_o, \vv k_o)$
has $\vv k_o$ normal to the orbit and $\vvj_o$ along the node of
the equatorial plane; the equatorial basis ${\cal B}_e = (\vv \imath_e,
\vvj_e, \vv k_e)$ has $\vvj_e=\vvj_o$ and $\vv k_e$
along the planet spin; The planet basis ${\cal B}_p = (\vv \imath_p,
\vvj_p, \vv k_p)$ has fixed vectors in the planet frame with $\vv k_p
= \vv k_e$.
}
\end{center}
\end{figure}

In this section, we make the approximation $\vv \ell = C \vv \omega$ in
such a way that the rotation vector is easily derived from the torque.
In the averaging process over the mean anomaly $M$ of the planet, the
orbit is Keplerian by definition and the angular momentum, as well as
the rotation vector, are fixed as they do not depend on $M$.  At this
stage, we shall introduce basis vectors which are used to compute the
secular torque. They are represented in Figure~\ref{fig.frame}. On the
one hand, the orbital motion is written in an orbital coordinate system
${\cal B}_o = (\vv \imath_o, \vvj_o, \vv k_o)$ such that $\vvj_o$ 
coincides with the ascending node of the equatorial plane and
$\vv k_o$ is normal to the orbit. On the other hand, the torque is
decomposed in an equatorial basis ${\cal B}_e = (\vv \imath_e, \vvj_e,
\vv k_e)$ constructed such that $\vvj_e = \vvj_o$
also points towards the node of the equator and $\vv k_e$ is along the
spin axis $\vv \omega$. The rotation angle between these two coordinate
systems is the obliquity denoted by $\theta$. For completeness,
Figure~\ref{fig.frame} also displays the basis ${\cal B}_p = (\vv
\imath_p, \vvj_p, \vv k_p)$ associated to the planet frame ${\cal
F}_p$ which differs from ${\cal B}_e$ by a rotation around $\vv k_e =
\vv k_p$. It should be stressed that even though the basis vectors of
${\cal B}_o$ and ${\cal B}_e$ are assumed constant during a revolution
period, they are not vectors of the inertial frame ${\cal F}_0$ because
both the planet and the orbit are precessing on long timescales. But
nothing prohibits to decompose an inertial vector in a non-inertial
coordinate system.

Let us introduce a few additional notations. We denote by $\vec x^o$ and
by $\vec x^e$ the coordinates of any vector $\vv x$ (computed with
respect to the inertial frame) in the bases ${\cal B}_o$ and ${\cal
B}_e$, respectively. We also define irregular solid
harmonics $S_{l,m}$ as
$$
S_{l,m}(\vec x) = \frac{1}{x^{l+1}} Y_{l,m}(\hvec x)\ .
$$
Solid harmonics transform in the same way as spherical harmonics under
rotation, thus
\begin{equation}
S_{l,m}(\vec x^e) = \sum_{m'=-l}^l \tens d^l_{m',m}(\theta) S_{l,m'}(\vec
x^o)\ ,
\label{eq.Slmrot}
\end{equation}
where $\tens d^l_{m',m}(\theta)$ is Wigner's d matrix
(Appendix~\ref{sec.Wigner}). The Keplerian elements used in the
following are the semi-major axis $a$, the mean motion rate $n$, the
eccentricity $e$, the true anomaly $v$, the mean anomaly $M$, and the
longitude of periastron $\varpi$ whose origin is the vector $\vv
\imath_0$. We denote by $X_k^{n,m}$ the Hansen coefficients defined such
that
$$
\left(\frac{r}{a}\right)^n \e^{\ii m v} = \sum_{k=-\infty}^\infty
X_k^{n,m} \e^{\ii k M}\ .
$$
Hansen coefficients are functions of eccentricity but this dependency is
dropped to simplify the notation. At least, we decompose the Fourier
transform of the Love distributions into their real and imaginary parts as
$$
\ubar k_l (\nu) = k_l^0 a_l(\nu) + \ii k_l^0 b_l (\nu) \ .
$$
With Maxwell rheology, we have
$$
a_l(\nu) = \frac{1+\tau_e\tau_l\nu^2}{1+\tau_l^2\nu^2} \ ,
\qquad
b_l(\nu) = -\left(1-\frac{\tau_e}{\tau_l}\right) \frac{\tau_l \nu}{1+\tau_l^2 \nu^2}\ .
$$

\subsection{Gravitational field coefficients}
In the body frame {\em and}\, in the frequency domain, gravitational field
coefficients are related to the external potential through \citep[e.g.,][]{Lambeck_book_1988}
$$
\ubar Z_{l,m}(\nu) = \ubar k_l(\nu)  \ubar Z^\star_{l,m}(\nu)\ .
$$
This relation expressed in the inertial equatorial frame  becomes (see
Appendix~\ref{sec.Fourier})
$$
\ubar z_{l,m}(\nu) = \ubar k_l(\nu - m \omega)\, \ubar z^\star_{l,m}(\nu)\ .
$$
Or, using the decomposition of the Fourier transform of the Love
distribution $k_l$,
\begin{equation}
\ubar z_{l,m}(\nu) = \Big(a_l(\nu-m\omega) + \ii b_l(\nu-m\omega)\Big) \ubar z^e_{l,m}(\nu)\ .
\label{eq.FTzlm}
\end{equation}
We now express the Fourier transform of $z^e_{l,m}(t)$. From its
definition (Eq.~\ref{eq.zedef}), we have
\begin{equation}
z^e_{l,m} (t) = -\delta_{l,2}\delta_{m,0} k_2^0 \frac{\omega^2R^3}{3Gm_1} + k_l^0 \frac{m_0}{m_1} R^{l+1} S_{l,m}(\vec x^e)\ .
\label{eq.zet}
\end{equation}
As said before, it is more simple to express solid harmonics in the
orbital frame. Indeed, in the latter frame, the colatitude of the radius
vector is $\pi/2$ and its longitude is $\varpi+v$. Thus, using the
expression of the spherical harmonics recalled in the
Appendix~\ref{sec.sh}, we get
\begin{align}
S_{l,m} (\vec x^o) &= (-1)^m \sqrt{\frac{(l-m)!}{(l+m)!}} P_{l,m}(0)
\frac{1}{x^{l+1}}\e^{\ii m (\varpi+v)} \nonumber \\
&= Y_{l,m}(\hvec \imath) \frac{\e^{\ii m \varpi}}{a^{l+1}}
\sum_{k=-\infty}^\infty X_k^{-(l+1),m} \e^{\ii k M}\ ,
\label{eq.Slmxo}
\end{align}
where $\hvec \imath$ is the unit vector of coordinates $(\hat \imath_+,
\hat \imath_0) = (-1/\sqrt 2, 0)$.  The Fourier transform of the
steady-state gravity coefficients $z_{l,m}$ in the inertial equatorial
frame is then deduced from Eqs.~(\ref{eq.Slmrot}), (\ref{eq.FTzlm}),
(\ref{eq.zet}), and (\ref{eq.Slmxo}). The result only contains terms at
frequencies $\nu_p = pn$, $p\in\mathbb{Z}$, which are given by
\begin{equation*}
\begin{split}
\ubar z_{l,m}(pn) =& -\delta_{l,2}\delta_{m,0}\delta_{p,0} a_2(0) k_2^0
\frac{\omega^2R^3}{3Gm_1} 
\\ & 
+ \ubar k_l(pn-m\omega)
\frac{m_0}{m_1} \left(\frac{R}{a}\right)^{l+1}
\sum_{m'=-l}^l \tens d^l_{m',m}(\theta) Y_{l,m'}(\hvec \imath)
X_p^{-(l+1),m'} \e^{\ii m' \varpi}\ .
\end{split}
\end{equation*}

\subsection{Secular torque}
The torque $\vec t = \dot{\vec \ell}$ (Eq.~\ref{eq.ldot}) involves the
angular operator $\vec J = (J_+, J_0, J_-)$. Let us denote by
$J^\mu_{l,m}$, $(\mu=+1,0,-1)$ the coefficient such that
$$
J_\mu (S_{l,m}(\vec x)) = J^\mu_{l,m} S_{l,m+\mu} (\vec x)\ .
$$
From the Appendix~\ref{sec.lo}, we have
$$
J^+_{l,m} = -\sqrt{\frac{l(l+1)-m(m+1)}{2}} 
\qquad \text{and} \qquad
J^0_{l,m} = m\ .
$$
With this notation,
$$
\dot \ell_\mu = -\ii \frac{Gm_0m_1}{R} \sum_{l=2}^{l_\mathrm{max}}
R^{l+1} \sum_{m=-l}^l \bar z_{l,m} J^\mu_{l,m} S_{l,m+\mu}(\vec x^e)\ .
$$
We expand $S_{l,m}(\vec x^e) = \sum_{m'} d^l_{m,m'}(\theta)
S_{l,m'}(\vec x^o)$ as above. Then, we substitute the steady-state
solution of $z_{l,m}$ previously found to get the steady-state torque
\begin{equation*}
\begin{split}
\dot \ell_\mu =& -\ii \frac{Gm_0m_1}{R} \sum_{l=2}^{l_\mathrm{max}}
R^{l+1} \sum_{m=-l}^l\sum_{m'=-l}^l\sum_{m"=-l}^l 
\sum_{p=-\infty}^\infty \sum_{p'=-\infty}^\infty \bar{\ubar k}_l(pn-m\omega) 
\\ & \times 
\bigg(-\delta_{l,2}\delta_{m,0}\delta_{p,0}\frac{\omega^2R^3}{3Gm_1} + 
\frac{m_0}{m_1}R^{l+1} d^l_{m',m}(\theta)\bar{\ubar S}_{l,m'}(pn)
\bigg) J^\mu_{l,m}
\\ & \times
d^l_{m",m+\mu}(\theta) \ubar S_{l,m"}(p'n)
\e^{\ii (p'-p)M}\ .
\end{split}
\end{equation*}
In this expression, $\ubar S_{l,m}(\nu)$ is the Fourier transform of
$S_{l,m}(\vec x^o(t))$ evaluated at the frequency $\nu$. The secular torque
is obtained for $p=p'$. The result is a function of $(\theta, a, e,
\varpi)$ and of the physical parameters of the problem, but it can be
simplified considering the fact that the pericenter is circulating
rapidly. We recall that $\ubar S_{l,m}$ is proportional to $\exp(\ii m
\varpi)$. Thus, $\langle \bar{\ubar S}_{l,m'} \ubar S_{l,m"} \rangle_\varpi$
is not zero only if $m'=m"$ and $\langle J_2 \ubar S_{l,m"} \rangle_\varpi
\neq 0$ when $m"=0$. The torque is further simplified by the symmetry of
the Love distributions, viz. $\ubar k_l(-\nu) = \bar{\ubar k}_l(\nu)$, or
equivalently, $a_l(-\nu)=a_l(\nu)$ and $b_l(-\nu)=-b_l(\nu)$. 
The average torque becomes
\begin{equation}
\begin{split}
\langle \dot \ell_\mu \rangle_{M,\varpi} = &
\frac{\ii}{3} k_2^0 m_0 \omega^2 R^5 
a_2(0) J^\mu_{2,0}
d^2_{0,\mu}(\theta) \ubar S_{2,0}(0)
\\ & - \ii \frac{Gm_0^2}{R} \sum_{l=2}^{l_\mathrm{max}}
R^{2l+2} \sum_{m=-l}^l \sum_{m'=-l}^l \sum_{p=-\infty}^\infty \ubar k_l(m\omega-pn)
\\ & \times
d^l_{m',m}(\theta) 
d^l_{m',m+\mu}(\theta)
J^\mu_{l,m}
|\ubar S_{l,m'}(pn)|^2
\end{split}
\label{eq.ldotmu}
\end{equation}

We now focus on the component $\mu=0$ of the secular torque which is
directly related to the evolution of the spin rate $\dot \omega$. Given
that $J^0_{l,m} = m$, terms in factor of $k_2^0$ disappear. Furthermore,
the term $T_{m,m',p} = m \ubar k_l(m\omega-pn) |d^l_{m',m}(\theta) 
\ubar S_{l,m'}(pn)|^2$ in the triple sum has the following symmetry 
$T_{-m,-m',-p} = -\bar T_{m,m',p}$. As a result,
\begin{equation*}
\langle \dot \ell_0 \rangle_{M,\varpi} = 2\frac{G m_0^2}{R}
\sum_{l=2}^{l_\mathrm{max}} R^{2l+2} k_l^0 \sum_{m=1}^{l} \sum_{m'=-l}^l
\sum_{p=-\infty}^\infty m b_l(m\omega-pn) 
|d^l_{m',m}(\theta) \ubar S_{l,m'}(pn)|^2\ .
\end{equation*}
Finally, we substitute the expression of $\ubar S_{l,m'}(pn)$ and we get
\begin{equation}
\begin{split}
\langle \dot \ell_0 \rangle_{M,\varpi} =& 2\frac{G m_0^2}{R}
\sum_{l=2}^{l_\mathrm{max}} \left(\frac{R}{a}\right)^{2l+2} k_l^0
\sum_{m=1}^{l} \sum_{m'=-l}^l \sum_{p=-\infty}^\infty m b_l(m\omega-pn)
\\ & \times
\left|d^l_{m',m}(\theta)Y_{l,m}(\hvec \imath) X_p^{-(l+1),m'}\right|^2\ .
\end{split}
\label{eq.l0dotgen}
\end{equation}
Note that in this sum, $m'$ is incremented by step of 2 because $m'$
should have the same parity as $l$ for $P_{l,m'}(0)\neq 0$ in the
expression of $Y_{l,m'}(\hvec \imath)$. At the
quadrupole order $l_\mathrm{max}=2$, the explicit expression is
\begin{equation}
\begin{split}
\langle \dot \ell_0 \rangle_{M,\varpi} =& k_2^0 \frac{Gm_0^2 R^5}{a^6} \sum_{k=-\infty}^\infty \Bigg( \\
& \quad b_2(2\omega-kn) (X_k^{-3,2})^2  \frac{3}{32} (1+\cos\theta)^4              \\
& +     b_2(2\omega-kn) (X_k^{-3,0})^2  \frac{3}{ 8} \sin^4\theta                  \\
& +     b_2(2\omega-kn) (X_k^{-3,-2})^2 \frac{3}{32} (1-\cos\theta)^4              \\
& +     b_2(\omega-kn)  (X_k^{-3,2})^2  \frac{3}{16} \sin^2\theta(1+\cos\theta)^2  \\
& +     b_2(\omega-kn)  (X_k^{-3,0})^2  \frac{3}{4}  \sin^2\theta\cos^2\theta      \\
& +     b_2(\omega-kn)  (X_k^{-3,-2})^2 \frac{3}{16} \sin^2\theta(1-\cos\theta)^2 
\Bigg)\ .
\end{split}
\label{eq.l0dot}
\end{equation}

The orthogonal component of the torque $\dot \ell_+$ does not present as
much symmetries as $\dot\ell_0$. From the general expression of $\langle\dot
\ell_\mu\rangle_{M,\varpi}$ (Eq.~\ref{eq.ldotmu}), we get
\begin{equation}
\begin{split}
\langle \dot \ell_+ \rangle_{M,\varpi} = &
\frac{\ii}{2\sqrt 2} \frac{k_2^0 m_0 \omega^2 R^5}{a^3(1-e^2)^{3/2}}\sin\theta\cos\theta
\\ & + \ii \frac{Gm_0^2}{R} \sum_{l=2}^{l_\mathrm{max}}
\left(\frac{R}{a}\right)^{2l+2} 
\sum_{m=-l}^l \sum_{m'=-l}^l \sum_{p=-\infty}^\infty
\sqrt{\frac{l(l+1)-m(m+1)}{2}} 
\\ & \times
\ubar k_l(m\omega-pn)
d^l_{m',m}(\theta) d^l_{m',m+1}(\theta)
\left|Y_{l,m'}(\hvec \imath)X_p^{-(l+1),m'}\right|^2\ .
\end{split}
\label{eq.lpdotgen}%
\end{equation}
At the quadrupole order and using the symmetries, this gives
\begin{equation}
\begin{split}
\langle \dot \ell_+ \rangle_{M,\varpi} = &
\frac{\ii}{2\sqrt 2} \frac{k_2^0 m_0 \omega^2 R^5}{a^3(1-e^2)^{3/2}}\sin\theta\cos\theta
\\ & + \frac{3\ii}{32\sqrt{2}}\frac{Gm_0^2R^5}{a^6}\sin\theta \sum_{p=-\infty}^\infty \Bigg(
\\ & + \big( \ubar k_2(  \omega-pn) -\bar{\ubar k}_2( 2\omega-pn) \big)
     \times \bigg(\left(X_p^{-3,-2}\right)^2 (1-\cos\theta)^3          
\\ & + 4\left(X_p^{-3, 0}\right)^2 \cos\theta\sin^2\theta               
     - \left(X_p^{-3, 2}\right)^2 (1+\cos\theta)^3\bigg)
\\ & + \big( \ubar k_2(        -pn) -\bar{\ubar k}_2(  \omega-pn) \big)
     \times \bigg(3 \left(X_p^{-3,-2}\right)^2\sin^2\theta(1-\cos\theta)                   
\\ & +4 \left(X_p^{-3, 0}\right)^2(3\cos^2\theta-1)\cos\theta
     -3 \left(X_p^{-3, 2}\right)^2\sin^2\theta(1+\cos\theta) \bigg)
\Bigg)\ .
\end{split}
\label{eq.lpdot}%
\end{equation}
Equations~(\ref{eq.l0dot}) and (\ref{eq.lpdot}) are written in a
specific coordinate system, viz. the equatorial basis ${\cal B}_e$. For
more generality, we now express the result in a vectorial form.  Let
$\hvec s$ and $\hvec k$ be the coordinates of the unit spin vector and
of the unit orbit normal in ${\cal F}_0$, respectively, i.e., $\hvec s =
\hvec k_e = \hvec k_p$ and $\hvec k = \hvec k_o$. The torque can formally
be decomposed as follows
\begin{subequations}
\begin{equation*}
\langle \dot {\vec \ell} \rangle_{M,\varpi} = t_1 \hvec s + t_2 \hvec k + t_3 \hvec k\times\hvec s\ .
\end{equation*}
with
\begin{equation*}
\langle \dot \ell_0 \rangle_{M,\varpi} = t_1 + t_2 \cos\theta
\qquad\text{and}\qquad
\langle \dot \ell_+ \rangle_{M,\varpi} = \frac{\sin\theta}{\sqrt 2}(t_2
- \ii t_3)\ .
\end{equation*}
\label{eq.coord2vec}%
\end{subequations}
The explicit expressions of $t_1$, $t_2$, and $t_3$ are displayed in
Table~\ref{tab.sectorque}.

\begin{table}
\begin{center}
\caption{\label{tab.sectorque}Components of the secular torque $\langle \dot{\vec \ell}
\rangle_{M,\varpi} = t_1 \hvec s + t_2 \hvec k + t_3 \hvec k\times\hvec
s$.}
\begin{tabular}{r l} \hline \hline
$t_1$ =& $\displaystyle  \quad \frac{3}{32} k_2^0 \frac{Gm_0^2R^5}{a^6} \sum_{k=-\infty}^\infty \Bigg(
            b_2(2\omega-kn) 
       \bigg( (X_k^{-3,2})^2  (1+\cos\theta)^4            
$ \\ & $\displaystyle
        +  4  (X_k^{-3,0})^2  \sin^4\theta
        +     (X_k^{-3,-2})^2 (1-\cos\theta)^4 \bigg)
      +  2  b_2(\omega-kn)  
$ \\[0.7em] & $\displaystyle
\times \bigg( (X_k^{-3,2})^2  \sin^2\theta(1+\cos\theta)^2
      +  4  (X_k^{-3,0})^2  \sin^2\theta\cos^2\theta    
$ \\[0.7em] & $\displaystyle
      +     (X_k^{-3,-2})^2 \sin^2\theta(1-\cos\theta)^2 \bigg)
\Bigg)
-t_2\cos\theta
$ \\[1em]
$t_2$ =&  $\displaystyle -\frac{3}{32}k_2^0\frac{Gm_0^2R^5}{a^6} \sum_{k=-\infty}^\infty \Bigg(
$ \\ & $\displaystyle
     + \Big(b_2(  \omega-kn) + b_2(2\omega-kn)\Big)
\times \bigg(
     \left(X_k^{-3,-2}\right)^2 (1-\cos\theta)^3          
$ \\[0.7em] & $\displaystyle
   + 4\left(X_k^{-3, 0}\right)^2 \cos\theta\sin^2\theta
   -  \left(X_k^{-3, 2}\right)^2 (1+\cos\theta)^3                   
\bigg)
$ \\[0.7em] & $\displaystyle
     + \Big(b_2(        -kn) + b_2(  \omega-kn)\Big)
\times \bigg(
     3\left(X_k^{-3,-2}\right)^2 \sin^2\theta(1-\cos\theta)                   
$ \\[0.7em] & $\displaystyle
    +4\left(X_k^{-3, 0}\right)^2 (3\cos^2\theta-1)\cos\theta
    -3\left(X_k^{-3, 2}\right)^2 \sin^2\theta(1+\cos\theta)
\bigg)
\Bigg)
$ \\[1em]
$t_3$ =& $\displaystyle -\frac{k_2^0 m_0 \omega^2 R^5}{2a^3(1-e^2)^{3/2}}\cos\theta
     - \frac{3}{32}k_2^0\frac{Gm_0^2R^5}{a^6} \sum_{k=-\infty}^\infty \Bigg(
$ \\ & $\displaystyle
     + \Big(a_2(  \omega-kn) - a_2(2\omega-kn)\Big)
\times \bigg(
     \left(X_k^{-3,-2}\right)^2 (1-\cos\theta)^3          
$ \\[0.7em] & $\displaystyle
   + 4\left(X_k^{-3, 0}\right)^2 \cos\theta\sin^2\theta
   -  \left(X_k^{-3, 2}\right)^2 (1+\cos\theta)^3                   
\bigg)
$ \\[0.7em] & $\displaystyle
     + \Big(a_2(        -kn) - a_2(  \omega-kn)\Big)
\times \bigg(
     3\left(X_k^{-3,-2}\right)^2 \sin^2\theta(1-\cos\theta)                   
$ \\[0.7em] & $\displaystyle
    +4\left(X_k^{-3, 0}\right)^2 (3\cos^2\theta-1)\cos\theta
    -3\left(X_k^{-3, 2}\right)^2 \sin^2\theta(1+\cos\theta)
\bigg)
\Bigg)
$ \\ \hline
\end{tabular}
\end{center}
\end{table}

In summary, Eqs.~(\ref{eq.l0dotgen},\ref{eq.lpdotgen}) provide the
general expression of the secular torque in the equatorial coordinate
system of the inertial frame. This torque is written explicitly at the
quadrupole order in Eqs.~(\ref{eq.l0dot},\ref{eq.lpdot}) and in a
vectorial form in Tab.~\ref{tab.sectorque}. It must be stressed that
these formulas are not limited to Maxwell bodies and can be applied to
any rheologies. They are exact in eccentricity but they involve an
infinite sum which has to be truncated. This sum is associated to the 
Fourier expansion of the orbital motion.

\subsection{Quasi-circular orbit}
At zero eccentricity, Hansen coefficients are given by $X_k^{n,m} =
\delta_{k,m}$. With this hypothesis, we retrieve the expressions 
(22) and (23) obtained by \citet{Correia_Icarus_2003} which
correspond to $\langle \dot \ell_0 \rangle_{M,\varpi} = t_1 + t_2
\cos\theta$ and $\langle \vec
k \cdot \dot{\vec \ell} \rangle_{M,\varpi} = t_1\cos\theta + t_2$,
respectively\footnote{Our notation is very similar to that of \citet{Correia_Icarus_2003} and
\citet{Cunha_IJA_2015} but, in these papers, $b^\mathrm{g}(\nu)$ is defined as the
opposite of the imaginary part of the Love number $\ubar k_2(\nu)$. Thus,
$b^\mathrm{g}(\nu)$ is related to our $b_2(\nu)$ through the relation $b^\mathrm{g}(\nu) =
-b_2(\nu)$.}.

In the case of low eccentric orbits, Hansen coefficients can be expanded at
second order according to
$$
X_0^{-3,0} = 1+\frac{3}{2}e^2
\ ,\qquad
X_1^{-3,0} = \frac{3}{2}e
\ ,\qquad
X_2^{-3,0} = \frac{9}{4}e^2
\ ,
$$
and
$$
X_1^{-3,2} = -\frac{1}{2}e
\ ,\qquad
X_2^{-3,2} = 1-\frac{5}{2}e^2
\ ,\qquad
X_3^{-3,2} = \frac{7}{2}e
\ ,\qquad
X_4^{-3,2} = \frac{17}{2}e^2
\ .
$$
With these values, we retrieve the expressions (10) and (11) of
\citet{Cunha_IJA_2015} which also correspond to $t_1+t_2\cos\theta$ and
$t_1\cos\theta+t_2$, respectively.

\subsection{Linear regime}
For completeness, we provide the vectorial decomposition of the torque 
in the linear regime $\tau_2\nu\ll 1$, where
$$
\b k_2(\nu) =
k_2^0\left(1-\ii\left(1-\frac{\tau_e}{\tau_2}\right)\ii\tau_2\nu\right)\ .
$$
From the definition of the Hansen coefficients, we get (see Appendix~B of
\citet{Correia_AA_2014}),
$$
\sum_{k=-\infty}^\infty \left(X_k^{n,m}\right)^2 = X_0^{2n,0}
\qquad\text{and}\qquad
\sum_{k=-\infty}^\infty k\left(X_k^{n,m}\right)^2 = m\sqrt{1-e^2}
X_0^{2n-2,0}\ .
$$
Substituting these equalities in the expressions of the
Table~\ref{tab.sectorque}, we recover the secular torque, Eqs.~(10,29)
of \citet{Correia_CeMDA_2011}, viz.
$$
\langle \dot {\vec \ell} \rangle_{M,\varpi} = 
- K
\tau_2 n \left(f_1(e)\frac{\hvec s + \cos\theta\hvec k}{2}\frac{\omega}{n} - f_2(e)\hvec k\right)
-\alpha \cos\theta\, \hvec k\times\hvec s\ ,
$$
with
\begin{align*}
K      &= \frac{3Gm_0^2R^5}{a^6} k_2^0 \left(1-\frac{\tau_e}{\tau_2}\right)\ , \\
\alpha &= \frac{1}{2}\frac{k_2^0 m_0 \omega^2 R^5}{a^3(1-e^2)^{3/2}}\ , \\
f_1(e) &= X_0^{-6,0} = \frac{1+3e^2+\frac{3}{8}e^4}{(1-e^2)^{9/2}}\ , \\
f_2(e) &= \sqrt{1-e^2} X_0^{-8,0} = 
\frac{1+\frac{15}{2}e^2+\frac{45}{8}e^4+\frac{5}{16}e^6}{(1-e^2)^6}\ .
\end{align*}

\subsection{Spin-rate and obliquity}
\label{sec.spinobli}
Let us assume that the orbit has most of the angular momentum of the
system. In that case, the equations of motion of the spin-rate and of the
obliquity are simply deduced from the secular torque
(Tab.~\ref{tab.sectorque}). One gets
\begin{equation}
\frac{1}{n}\frac{d\omega}{dt} = \frac{t_1 + t_2\cos\theta}{Cn}
\qquad \text{and} \qquad
\frac{d\theta}{dt} = -\frac{t_2\sin\theta}{C\omega}\ .
\label{eq.omi}
\end{equation}
\begin{figure}
\begin{center}
\includegraphics[width=\linewidth]{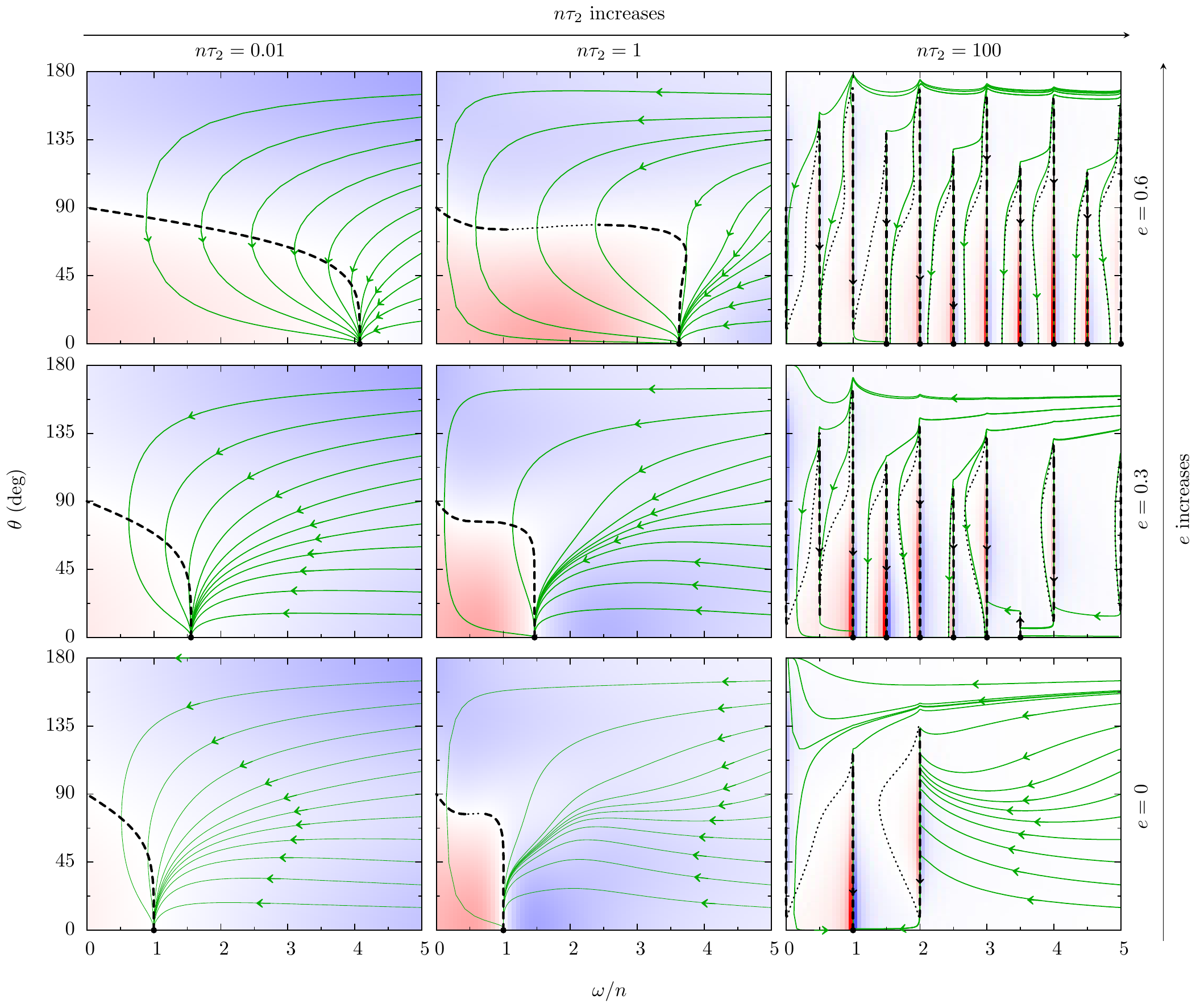}
\caption{\label{fig.field}Secular trajectories of the spin-axis in the
plane obliquity $\theta$ versus rotation $\omega/n$. From left to right,
the product $n\tau_2$ increases from 0.01 to 100. From bottom to top,
the eccentricity increases from 0.0 to 0.6. Trajectories of the
spin-axis are represented by green solid curves, the background color
represents the derivative of $\omega$: in blue $\omega$ decreases, in
red $\omega$ increases. The locus of rotation equilibria such that
$d\omega/dt=0$ (irregardless of $d\theta/dt$) are highlighted by black
curves. Dashed ones and dotted ones represent stable and unstable
equilibria, respectively. Black dots on the $x$-axis are the fixed
points. This figure has been made by integration of
Eqs.~(\ref{eq.omi}) with $t_1$, $t_2$, and $t_3$ taken from
Tab.~\ref{tab.sectorque}. Sums have been truncated at $|k|\leq 100$.}
\end{center}
\end{figure}%
It should be noted that the trajectory of the spin in the plane
$(\omega/n, \theta)$ only depends on the ratio $\omega/n$, the obliquity
$\theta$, the eccentricity $e$, and the product $n\tau_2$. A few of them
are plotted in Figure~\ref{fig.field} for $e\in\{0,0.3,0.6\}$ and
$n\tau_2\in\{0.01,1,100\}$. Plots are limited to positive $\omega/n$ but
they can be extended to negative rotations with the symmetry
$ (\omega, \theta) \leftrightarrow (-\omega, \pi-\theta)\ . $ Indeed,
these two pairs are equivalent although they do not correspond to the
same physical state \citep{Correia_Laskar_nature_2001}.

For $n\tau_2 = 0.01$ (Fig.~\ref{fig.field} left column), i.e. when the
viscous timescale is much shorter than the orbital period, the system is
in the linear regime. All trajectories converge smoothly towards a
prograde pseudo-synchronous rotation on the $x$-axis. Evolutions are
free from temporary captures in spin-orbit resonance.

At $n\tau_2 = 100$ (Fig.~\ref{fig.field} right column), the viscous
timescale is much greater than the orbital period. Resonant features
appear in the phase space even at zero eccentricity. Indeed, when $e=0$,
if the planet is tilted, its rotation can be trapped in three different
spin-orbit resonances, namely the 0:1, the 1:1, and the 2:1. However, the
obliquity is decreasing along these resonances and, in the planar
configuration, only the synchronous holds. The final state is thus
the synchronous rotation. At higher eccentricities, we observe many more
spin-orbit equilibria for which $\omega/n$ is a half integer. As in the
circular case, a few of these resonances disappear at zero obliquity but
several do persist. The case $e=0.3$ shows an interesting feature: let
us consider a trajectory (not represented) starting at $\omega/n=4.5$ and
$\theta=90^\circ$. Because this point is in a blue region, $\omega$
decreases until the rotation reaches the 4:1 resonance. Then,
the system follows the resonance downward until the obliquity reaches
about $5^\circ$ where the resonance disappears. The subsequent evolution
is horizontal toward the 7:2 resonance. But this resonance is special
because $d\theta/dt>0$. Thus, the system climbs this resonance up to its
end at $\theta\approx16^\circ$. The field line continues on the left
towards the 3:1 spin-orbit resonance. At last, this resonance has a
``normal'' behavior, the obliquity decreases and the system ends up in
a planar state with $\omega/n = 3$. This prediction has been tested
numerically by integration of the instantaneous equations of motion
(Eqs.~\ref{eq.evolz}) (see Section~\ref{sec.inst_vs_sec}). At higher
eccentricity ($e=0.6$), all spin-orbit resonances displayed in
Fig.~\ref{fig.field}, i.e. with $\omega/n\leq 5$, are such that
$d\theta/dt$ is negative.  Thus, along these resonances the obliquity
varies in a monotonous way toward the planar configuration. Note that
if a system starts with a fast rotation $\omega \gg n$, it will almost
certainly never reach an intermediate spin-orbit resonance such as the
2:1 or the 3:1 because the obliquity would have to be very fine tuned
close around $164^\circ$ at $\omega/n=5$. 

For $n\tau_2=1$ (Fig.~\ref{fig.field} middle column), the evolution does
not show any spin-orbit resonances. The phase space is qualitatively
similar to that of the linear regime. Field lines are only slightly
deformed.

\section{Application to HD~80606\,b}
\label{sec:results}
In this section, we apply the model at the quadrupole order
$l_\mathrm{max}=2$ to HD~80606\,b. The formalism is the same as in
\citep{Correia_AA_2014}, except that only the planar case was studied in
this previous work. Here, we extend the analysis to the spatial case by
allowing non-zero initial obliquities. First, we briefly recall the
results obtained for HD~80606\,b in the planar case, with $\tau_e=0$ and
$\tau_2$ ranging between $10^{-5}$ and $10^0$ yr. Then, we present our
results in the spatial problem.

\subsection{Description of the planar evolution}
As shown by the differential equation (\ref{eq.zze}), tides can be seen
as a low-pass filter between the excitation $Z^e_{l,m}$ and the
response $Z^\nu_{l,m} = Z_{l,m}$. If the cutoff frequency
$1/\tau_2$ is much greater than the orbital frequency $n$, i.e., $\tau_2
\ll 10^{-2}$ yr, all the ``signal'' is transmitted by the filter but
with a small phase shift. This is equivalent to the constant time-lag
model $\Delta t = \tau_2$. The surface of the planet undergoes strong
deformations at the orbital frequency but the amount of dissipation is
low because of the weak viscosity. Once the spin of the planet is
damped, it follows a pseudo-synchronous equilibrium $\Omega_e$ which is a
function of the eccentricity $e$. Here, we recall its expression in the
spatial case, i.e. with obliquity $\theta$, in anticipation to the
forthcoming section. We have \citep[e.g.,][]{Correia_CeMDA_2011}
\begin{equation}
\frac{\Omega_e}{n} = \frac{1+\frac{15}{2}e^2+\frac{45}{8}e^4+\frac{5}{16}e^6}
                          {(1-e^2)^{3/2}\left(1+3e^2+\frac{3}{8}e^4\right)}
                     \frac{2\cos\theta}{1+\cos^2\theta}\ .
\label{eq.Ome}
\end{equation}

For $\tau_2 \gg 10^{-2}$ yr, the cutoff frequency is less than the mean
motion rate. The deformation of the planet, represented by
$$
J_2=-Z_{2,0} \qquad\mathrm{and}\qquad 
\epsilon=\sqrt{C_{22}^2 +S_{22}^2} = \sqrt{6}|Z_{22}|\ ,
$$
only {\em sees} a mean excitation averaged over the mean anomaly and
takes the expression
\citep{Correia_AA_2014}
\begin{subequations}
\begin{align}
\langle J_2 \rangle_M &= k^0_2 \left(\frac{\Omega^2 R^3}{3Gm_1}
+
\frac{1}{2}\frac{m_0}{m_1}\left(\frac{R}{a}\right)^3(1-e^2)^{3/2}\right)\
, 
\label{eq.J2e} \\
\langle \epsilon_p \rangle_M &= \frac{k^0_2}{4} \frac{m_0}{m_1}
\left(\frac{R}{a}\right)^3 X^{-3,2}_{2p}(e)\ ,
\label{eq.epse}
\end{align}%
\label{eq.deformeq}%
\end{subequations}
with $p=[2\Omega/n]/2$, where $[x]$ means the nearest integer of $x$
($[x]\in\mathbb{Z}$ and $[x]-1/2\leq x < [x]+1/2$).  Thus,
despite a high viscosity, dissipation is low because the deformation is
weak and slow.  In that case, the constant time-lag model does not hold
anymore. The planet rotation gets trapped in spin-orbit resonances
$\Omega/n=p$, the pseudo-synchronous state is not an equilibrium
anymore.

At $\tau_2 \approx 10^{-2}$ yr, the orbital frequency is of the same
order of magnitude as the cutoff frequency. A few harmonics of the
orbital period pass the filter and are retrieved in the deformation
of the planet. Moreover, the viscosity is higher than in the
constant-time lag regime. Both effects generate strong dissipation and
a fast decay of the semi-major axis and eccentricity. 

\subsection{Fast damping of the obliquity and subsequent planar
evolution}
Numerical simulations were performed using the formalism in the inertial
reference frame (Sect.~\ref{sec.F0}, Eqs.~\ref{eq.evolz}). We have
tested different values of $\tau_2$, but the main conclusion of this
section remains unchanged. Thus, we only present results corresponding
to the intermediate case $\tau_2=10^{-2}$ yr.

\begin{figure}[h!]
\begin{center}
\includegraphics[width=0.5\linewidth]{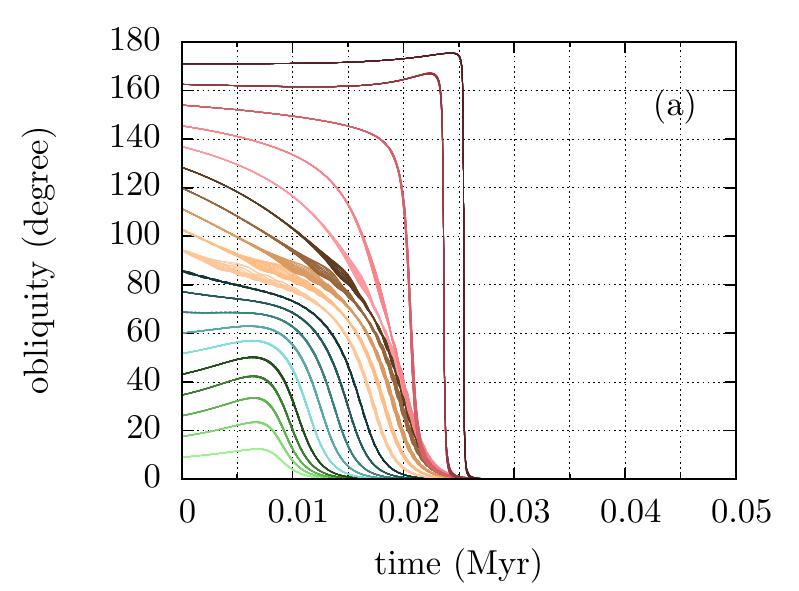}%
\includegraphics[width=0.5\linewidth]{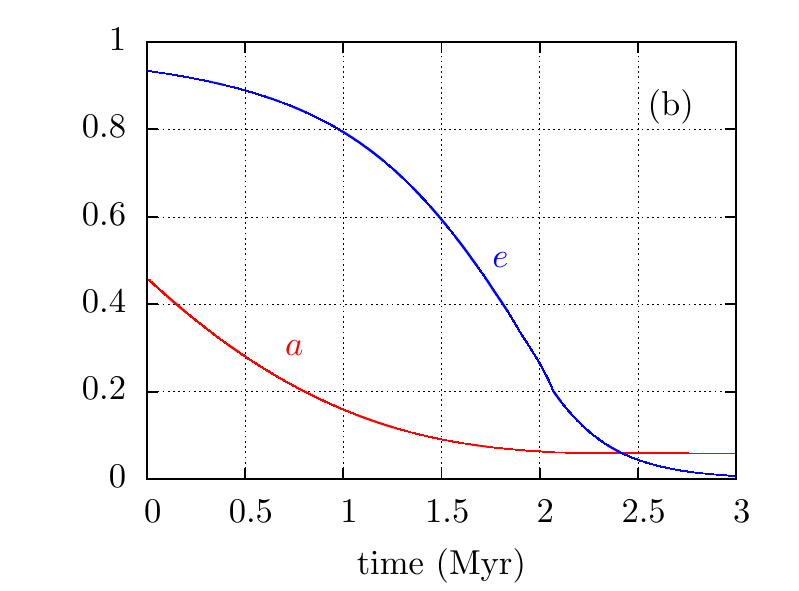}
\includegraphics[width=0.5\linewidth]{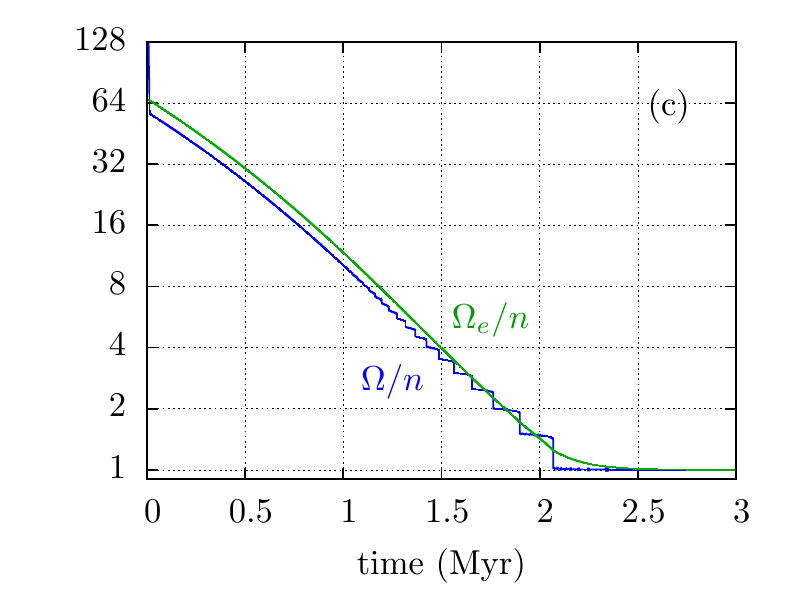}%
\includegraphics[width=0.5\linewidth]{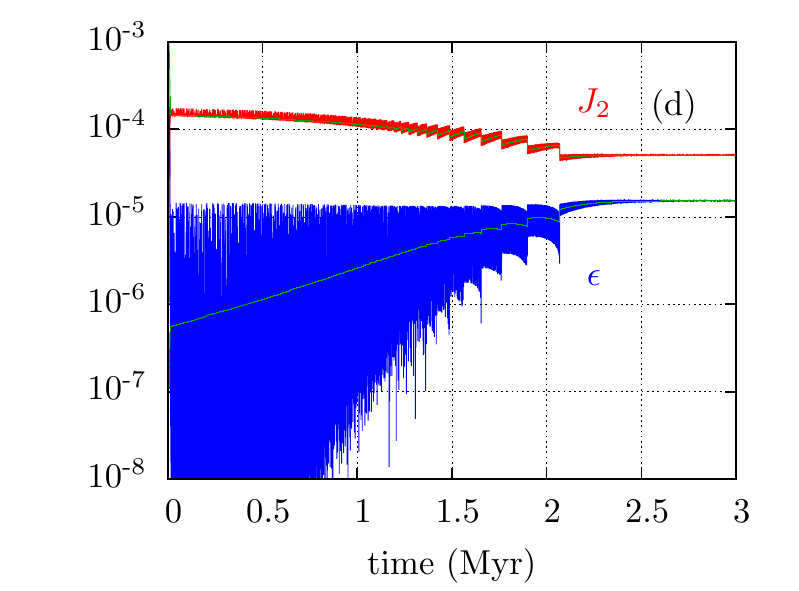}
\caption{\label{fig.hd80606}Time evolution of HD~80606\,b for
$\tau_2=10^{-2}$ yr ($0.21 \leq n\tau_2 \leq 0.95$). Panel ({\em a}) shows 400
evolutions with a grid of initial spin-axes (20 obliquities ranging
between 0 and 180 degrees times 20 precession angles ranging between 0
and 360 degrees).  The other three figures are initialized with an
obliquity of 60 degrees and a precession angle of 0 degree. We plot the
obliquity ({\em a}), the semi-major axis (in au) and the eccentricity
({\em b}), the ratio between the planet spin rate and the orbital mean
motion ({\em c}), and the planet $J_2$ and $\epsilon$ ({\em d}). The
green line gives the equilibrium rotation (Eq.~\ref{eq.Ome}) ({\em c}),
and the equilibrium values for $J_2$ and $\epsilon$, respectively
(Eqs.~\ref{eq.J2e} and \ref{eq.epse}) ({\em d}).}
\end{center}
\end{figure}

Figure~\ref{fig.hd80606}a depicts the evolution of the planet obliquity
for 400 different initial conditions: 20 obliquities regularly spread
between 0 and 180 degrees times 20 precession angles equispaced over 360
degrees. The initial precession angle does not play a significant role
in the evolution of the system.  At a given initial obliquity, all
integration's closely follow the same track.  This result strengthens the
approximation made in the previous section where we averaged the secular
equations of motion over the longitude of the pericenter $\varpi$. In
comparison, obliquities starting at different values can have distinct
initial slopes. But in all cases, the obliquity is fully damped before
30 kyr, a timescale much shorter than that of the orbital decay.

The subsequent evolution ($t>30$ kyr) is done at zero obliquity. The
problem is thus fully described by the planar model. Indeed, we
recover the results displayed in \citep[Fig.~6]{Correia_AA_2014}. The
semi-major axis and the eccentricity are damped within a timescale of 2
Myr (Fig.~\ref{fig.hd80606}b), the spin rate of the planet follows a
series of resonances with the orbital mean motion
(Fig.~\ref{fig.hd80606}c), and the deformation of the planet oscillates
with intermediate amplitudes around its equilibrium given by
Eqs.~\ref{eq.deformeq}.

Numerical experiments performed with different values of $\tau_2$ are
similar. Obliquities are fully damped in a timescale much shorter than
those associated to the semi-major axis and the eccentricity. Once the
system becomes planar, we retrieve the evolution observed in
\citep{Correia_AA_2014}. This result reveals that the motion of
the spin-axis can be followed independently from that of the orbit.
Thus, we can directly compare the numerical solutions of the instantaneous
equations of motion (Eqs.~\ref{eq.evolz}) to those dictated by the
secular torque (Section~\ref{sec.secular}).

\subsection{Instantaneous versus secular evolution}
\label{sec.inst_vs_sec}
In this section, we keep the system HD~80606\,b as a proxy to analyze
the spatial evolution of spin-axes given by the instantaneous equations
of motion (Eqs.~\ref{eq.evolz}). To start a simulation at a given
eccentricity $e$, we choose the semi-major axis $a$ as if the system had
evolved from its current orbit ($e_0=0.933$ and $a_0=0.455$ au) with a
constant angular momentum, i.e., such that $a(1-e^2) = a_0(1-e_0^2)$.
In all simulations, we set the initial precession angle and the initial 
longitude of periapsis to zero.

\begin{figure}[t]
\begin{center}
\includegraphics[width=\linewidth]{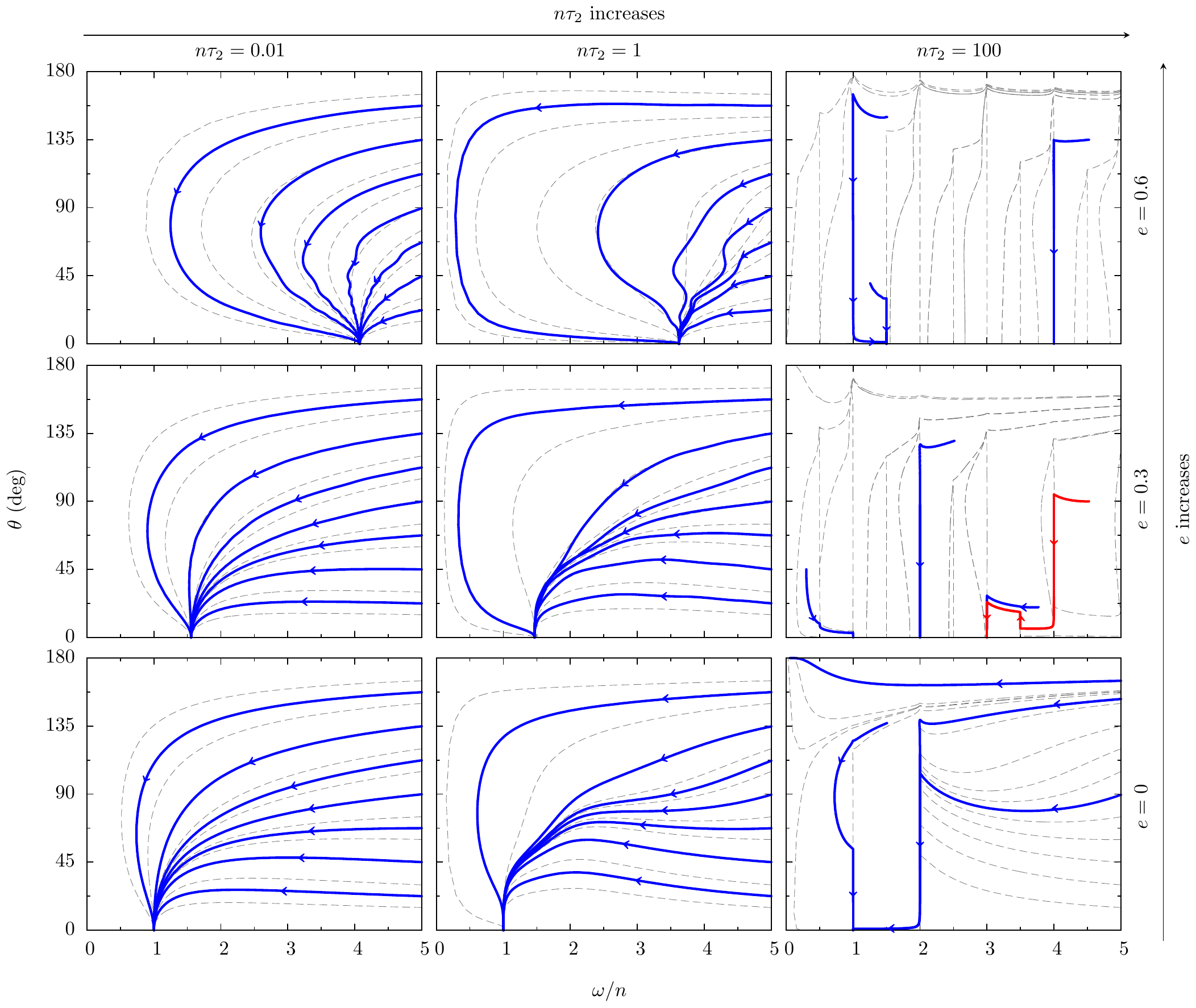}
\caption{\label{fig.simu}Instantaneous versus secular evolution. 
Solid thick curves are the trajectories obtained by integration of the
instantaneous equations of motion (Eqs.~\ref{eq.evolz}). Thin dashed
curves are the field lines of the Figure~\ref{fig.field} obtained by
integration of the secular equations of motion (Eqs.~\ref{eq.omi}). The
red curve is the trajectory discussed in Section~\ref{sec.spinobli}.}
\end{center}
\end{figure}

\begin{figure}
\begin{center}
\includegraphics[width=0.8\linewidth]{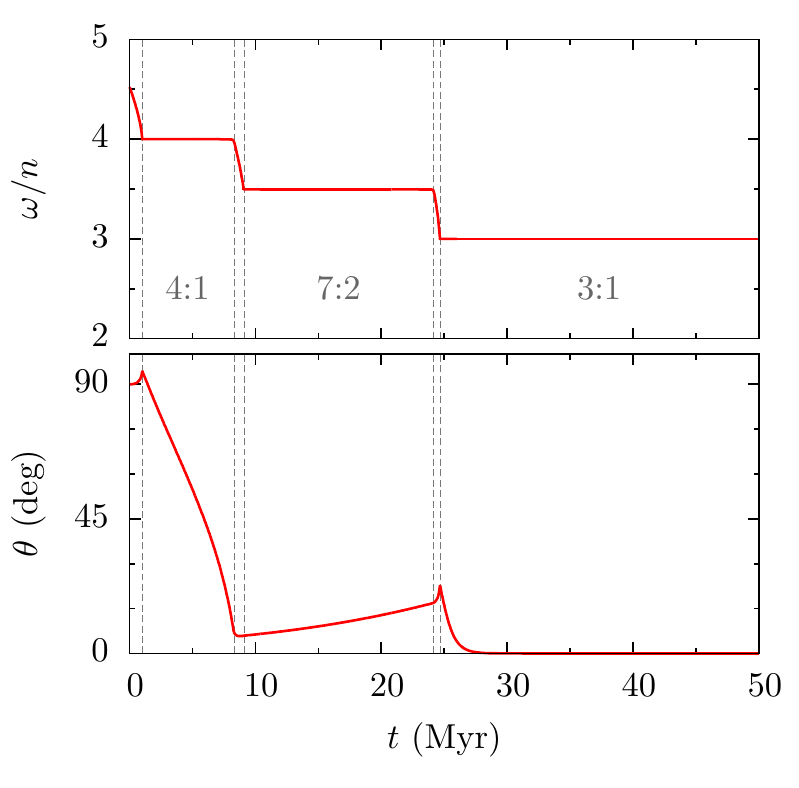}
\caption{\label{fig.trajectory}Time evolution of HD~80606\,b with the
same initial conditions as the red trajectory in
Figure~\ref{fig.simu} ($e=0.3$ and $n\tau_2=100$). 
Vertical dashed lines delimit regions of spin-orbit resonance. Note that
after 50 Myr, the eccentricity and the semi-major axis have only
decreased by about $3\times10^{-4}$ and $10^{-3}$ in relative value,
respectively. They can thus be considered constant as in
Figure~\ref{fig.simu}.}
\end{center}
\end{figure}

Figure~\ref{fig.simu} displays the results in the plane
$(\omega/n,\theta)$ as in Figure~\ref{fig.field} together with the
secular field lines obtained in Section~\ref{sec.secular}. The match
between the two approaches is excellent. Solutions of the instantaneous
equations of motion (Eqs.~\ref{eq.evolz}) closely follow the paths
dictated by the secular torque (Eqs.~\ref{eq.omi}), except however for
$e=0.6$ and $n\tau_2=1$ where the instantaneous evolutions exhibit more
wiggles than the secular ones. In particular, we retrieve the special
trajectory at $e=0.3$ and $n\tau_2=100$ discussed in
Section~\ref{sec.spinobli} which starts at $\omega/n=4.5$ and
$\theta=90^\circ$ (plotted in red in Figure~\ref{fig.simu}). The time
evolution of this trajectory is shown in Figure~\ref{fig.trajectory}.
We see that the system spends most of the time in spin-orbit resonant
configurations. The temporal evolution also emphasizes the peculiar
behavior of the 7:2 resonance in which the planet obliquity increases.

\section{Conclusion}
\label{sec:conclusion}
In this paper, we present a tidal theory based on the Maxwell rheology
valid in the spatial case. This extends the models presented by
\citet{Ferraz-Mello_CeMDA_2013} and \citet{Correia_AA_2014} which were
restricted to planar configurations. 

The evolution of the deformation of the planet, given by a first order
differential equation (Eqs.~\ref{eq.dZnu} and \ref{eq.dznu}), is integrated
numerically together with the orbital motion. As already noted by
\citet{Correia_AA_2014}, this way allows to compute the instantaneous
variation of the shape of the planet for all perturbations, even for
non-periodic ones. There is no need to decompose the excitation in an
infinite Fourier series as in, e.g., \citet{Kaula_RvGSP_1964}.
By consequence, the formalism is regular at all eccentricities,
spin rates, and obliquities.

For this problem, we have chosen a formalism taken from quantum theory,
conceived for angular momentum representations, and based on complex
spherical harmonics $Y_{l,m}$. Our choice has been motivated by the
following reasons: the gravitational potential of the planet is easily
expanded in $Y_{l,m}$; $Y_{l,m}$'s can be conveniently expressed in
terms of Cartesian coordinates; tidal force and torque have compact
expressions because $Y_{l,m}$'s are the eigenvectors of the ladder
operators $\vec \nabla$ and $\vec J=-\ii\vec x \times \vec \nabla$;
our model is given at any multipole order thanks to the recurrence
relations present in many quantum mechanics textbooks such as
\citep{Varshalovich_book_1988}.

Tidal equations are naturally written in the frame of the body, but this
choice is not convenient for the analysis of the orbital evolution.
Here, we provide the equations of motion both in the body frame ${\cal
F}_p$ (Eqs.~\ref{eq.evolZ}) and in the inertial frame ${\cal F}_0$
(Eqs.~\ref{eq.evolz}). If the planet does not have any permanent zonal
coefficients, the description of the problem in ${\cal F}_0$ presents a
numerical advantage. Indeed, whatever is the rotation speed of the
planet, the tidal bulge follows the perturbing body. Thus, the integration
time step can be adjusted to the orbital motion even if the planet
rotates much faster.

The equations of motion written in the inertial frame allowed us to
compute the secular tidal torque as a Fourier series averaged
over the orbital revolution and over the precession period. We provide
an explicit vectorial expression of this torque at the quadrupolar order
as well as the general expression at any multipole order. Maps of the
secular evolution of the spin-axis show many resonant features when the
viscous timescale is longer than the orbital period. This
characteristic was already present in planar studies but here we
observe that non-synchronous spin-orbit resonances appear in the spatial
case even at zero eccentricity. In most of these resonant states, the
obliquity decreases to zero, but we found a peculiar situation were the
obliquity is instead growing.

We applied our model to HD~80606 with different values of relaxation
time and different initial obliquities. We observed that in all cases,
the obliquity is damped faster than the semi-major axis and the
eccentricity. Once the system becomes planar, the evolution follows
the path described in \citep{Correia_AA_2014}. In particular, when the
relaxation time is greater than the orbital period, the planet gets
trapped in successive spin-orbit resonances even though it does not have
any permanent multipole \citep{Correia_AA_2014}. We have also analyzed
in more detail the evolution of the spin-axis during the phase where the
obliquity is not fully damped. Results are in good agreement with the
predictions made with the averaged equations. We nevertheless observe
wiggles at high eccentricity which were not present in the secular
phase-space. 

Our model can also be applied to close-in super-Earths for which the
relaxation time of the mantle is almost certainly longer than the
orbital period. As these planets are often found with planetary
companions, their eccentricities are never exactly zero
\citep[e.g.,][]{Laskar_AA_2012}. This implies that short-period
terrestrial exoplanets are likely in spin-orbit resonances
\citep{Correia_AA_2014}. 
In addition, as in the solar system, they also present small mutual
inclinations of about $1^\circ$ on average
\citep{Tremaine_Dong_AJ_2012,
Figueira_AA_2012, Fabrycky_ApJ_2014}. This value is large enough to
perturb the long-term evolution of their obliquity and, even if the
orbit is circular, a forced obliquity can trap the rotation in a
non-synchronous spin-orbit resonance state.  Our formalism is thus well
adapted to model the evolution of these planets spin-axis and to infer
constraints on their habitability. We also envision to extend
the formalism to thermal atmospheric tides which have the same frequency
dependence as Maxwell rheology \citep{Auclair_CeMDA_2016}.

\begin{acknowledgements}
GB is grateful to Dan Fabrycky for the fruitful discussions  which lead
to this work. We acknowledge support from CIDMA strategic project UID/MAT/04106/2013. 
\end{acknowledgements}

\appendix

\section{Spherical harmonic}
\label{sec.sh}
By convention, Legendre associated polynomials are defined as
\begin{equation}
P_{l,m}(x) = \frac{1}{2^l l!} (1-x^2)^{m/2}
\frac{d^{l+m}}{dx^{l+m}}(x^2-1)^l\ ,
\end{equation}
with the symmetry
\begin{equation}
P_{l,-m} (x) = (-1)^m \frac{(l-m)!}{(l+m)!}P_{l,m}(x)\ .
\end{equation}
The Schmidt semi-normalized spherical harmonics are defined as
\begin{equation}
Y_{l,m}(\theta,\phi) = (-1)^m \sqrt{\frac{(l-m)!}{(l+m)!}}
P_{l,m}(\cos \theta) \mathrm{e}^{\ii m\phi}
\end{equation}
with the symmetry
\begin{equation}
Y_{l,-m}(\theta,\phi) = (-1)^m \bar Y_{l,m}(\theta,\phi) \ .
\label{eq.symmetry}
\end{equation}
Using the complex Cartesian coordinate system as defined in
\citep{Varshalovich_book_1988}, for any unit vector $\hat x$, we have
\begin{subequations}
\begin{align}
Y_{0,0}(\hvec x)    &= 1 \ , \\
Y_{1,0}(\hvec x)    &= \hat x_0 \ , \\
Y_{1,1}(\hvec x)    &= \hat x_+ \ , \\
l\, Y_{l,0}(\hvec x) &= 
      (2l-1)\hat x_0 Y_{l-1,0}(\hvec x) - (l-1)Y_{l-2,0}(\hvec x)\ , 
\label{eq.rec1}
\\
\sqrt{l+m} Y_{l,m}(\hvec x) &= \sqrt{l-m}\, \hat x_0
Y_{l-1,m}(\hvec x) + \sqrt{2(l+m-1)} \hat x_+ Y_{l-1,m-1}\ .
\label{eq.rec2}
\end{align}
\end{subequations}
The last two equations (\ref{eq.rec1} and \ref{eq.rec2}) allow to
recursively compute all spherical harmonics of order $m\geq 0$. Those
with $m<0$ are deduced from the symmetry relation (\ref{eq.symmetry}).
Up to the degree 3 included, we have
   \begin{equation}
   \left\{
   \begin{array}{ll}
   Y_{2,0} &= \displaystyle \frac{1}{2}(3 \hat x_0^2 -1) \\[1em]
   Y_{2,1} &= \displaystyle \sqrt 3 \hat x_0 \hat x_+ \\[1em]
   Y_{2,2} &= \displaystyle \frac{\sqrt 6}{2} \hat x_+^2
   \end{array}
   \right.\ ,
   \qquad
   \left\{
   \begin{array}{ll}
   Y_{3,0} &= \displaystyle \frac{5}{2}\hat x_0^3 - \frac{3}{2}\hat x_0 \\[1em]
   Y_{3,1} &= \displaystyle \frac{\sqrt 6}{4} \left(5 \hat x_0^2 \hat x_+ - \hat x_+\right) \\[1em]
   Y_{3,2} &= \displaystyle \frac{\sqrt{30}}{2} \hat x_0 \hat x_+^2 \\[1em]
   Y_{3,3} &= \displaystyle \frac{\sqrt{10}}{2} \hat x_+^3
   \end{array}
   \right.\ .
   \end{equation}

\section{Ladder operators}
\label{sec.lo}
Regular solid harmonics $x^l Y_{l,m}(\hvec x)$ and irregular ones
$Y_{l,m}(\hvec x)/x^{l+1}$ are eigenvectors of each component
of the gradient operator $\vec \nabla = (\nabla_+, \nabla_0, \nabla_-)$
and of the angular momentum operator $\vec J = (J_+, J_0, J_-)$. The
respective eigenvalues can be found in \citep[e.g.,][]{Varshalovich_book_1988}.
We have
\begin{align}
\nabla_+ \left(x^{l}Y_{l,m}(\hvec x)\right) &=
-\sqrt{\frac{(l-m-1)(l-m)}{2}} \,x^{l-1}Y_{l-1,m+1}(\hvec x) \nonumber \\
\nabla_0 \left(x^{l}Y_{l,m}(\hvec x)\right) &=
+\sqrt{(l+m)(l-m)} \,x^{l-1} Y_{l-1,m}(\hvec x) \\
\nabla_- \left(x^{l}Y_{l,m}(\hvec x)\right) &=
-\sqrt{\frac{(l+m-1)(l+m)}{2}} \,x^{l-1}Y_{l-1,m-1}(\hvec x)\ , \nonumber
\end{align}
\begin{align}
\nabla_+ \left(\frac{1}{x^{l+1}}Y_{l,m}(\hvec x)\right) &=
-\sqrt{\frac{(l+m+1)(l+m+2)}{2}} \frac{1}{x^{l+2}}Y_{l+1,m+1}(\hvec x) \nonumber \\
\nabla_0 \left(\frac{1}{x^{l+1}}Y_{l,m}(\hvec x)\right) &=
-\sqrt{(l+m+1)(l-m+1)} \frac{1}{x^{l+2}} Y_{l+1,m}(\hvec x) \\
\nabla_- \left(\frac{1}{x^{l+1}}Y_{l,m}(\hvec x)\right) &=
-\sqrt{\frac{(l-m+1)(l-m+2)}{2}} \frac{1}{x^{l+2}}Y_{l+1,m-1}(\hvec x)\ , \nonumber
\end{align}
and
\begin{align}
J_+ \bigg( f(x) Y_{l,m}(\hvec x) \bigg) &= -\sqrt{\frac{l(l+1)-m(m+1)}{2}}
f(x) Y_{l,m+1}(\hvec x) \nonumber \\
J_0 \bigg( f(x) Y_{l,m}(\hvec x) \bigg) &= m f(x) Y_{l,m}(\hvec x) \\
J_- \bigg( f(x) Y_{l,m}(\hvec x) \bigg) &= +\sqrt{\frac{l(l+1)-m(m-1)}{2}}
f(x) Y_{l,m-1}(\hvec x)\ , \nonumber
\end{align}
where $f(x)$ is any function of the modulus $x=\|\vv x\|$.

\section{Rotation and Wigner matrices}
\label{sec.Wigner}
Let a vector $\vv x$ and two coordinate systems ${\cal B}$ and ${\cal B}'$ such that
$\vec x$ and $\vec x'$ are the coordinates of $\vv x$ in ${\cal B}$ and
${\cal B}'$, respectively. Let us further assume that $\vec x$ and $\vec
x'$ are related to each other by a rotation of the form
$$
\vec x = {\tens R}_3(\alpha) {\tens R}_2(\beta) {\tens R}_3(\gamma) \vec x'\ ,
$$
where ${\tens R}_3$ and ${\tens R}_2$ are the matrices of rotation around
the third and the second axes, respectively. Wigner D matrix 
$\tens D^l_{m,m'}(\alpha,\beta,\gamma)$ is defined such that
\citep[e.g.,][]{Varshalovich_book_1988}
\begin{equation}
Y_{l,m}(\hvec x') = \sum_{m'=-l}^l \tens D^l_{m',m}(\alpha,\beta,\gamma)
Y_{l,m'}(\hvec x)\ .
\label{eq.rotation}
\end{equation}
Each element $\tens D^l_{m,m'}(\alpha,\beta,\gamma)$ can be written as
\citep[e.g.,][]{Varshalovich_book_1988}
\begin{equation}
\tens D^l_{m,m'}(\alpha,\beta,\gamma) = \e^{-\ii m \alpha} \tens
d^l_{m,m'}(\beta) \e^{-\ii m'\gamma}\ ,
\label{eq.dsym}
\end{equation}
where $\tens d^l_{m,m'}(\beta)$ is the Wigner d matrix. The inverse
$\tens D^l_{m,m'}(-\gamma,-\beta,-\alpha)$ is given by the adjoint 
$\bar{\tens D}^l_{m',m}(\alpha,\beta,\gamma)$ of $\tens D^l_{m,m'}(\alpha,\beta,\gamma)$:
$$
\tens D^l_{m,m'}(-\gamma,-\beta,-\alpha) = \e^{\ii m'\alpha}
\tens d^l_{m',m}(\beta)\e^{\ii m\gamma}\ .
$$
The
convention 3-2-3 of the rotation (Eq.~\ref{eq.rotation}) is such that
$\tens d^l_{m,m'}(\beta)$ is a real function. Wigner d matrix possesses many
symmetries, among which \citep[e.g.,][]{Varshalovich_book_1988}
$$
\tens d^l_{m,m'}(\beta) 
  = (-1)^{m-m'}\tens d^l_{-m,-m'}(\beta)
  = (-1)^{m-m'}\tens d^l_{m',m}(\beta)
  = \tens d^l_{-m',-m}(\beta)\ .
$$
Wigner d matrix can be constructed recursively using the hereinabove
symmetries, the following initialization
\citep[e.g.,][]{Varshalovich_book_1988}
\begin{equation}
\tens d^0_{0,0}(\beta) = 1\,,\ 
\tens d^1_{0,0}(\beta) = \cos\beta \,,\ 
\tens d^1_{1,-1}(\beta) = \frac{1-\cos\beta}{2} \,,\ 
\tens d^1_{1,0}(\beta) = -\frac{\sin\beta}{\sqrt 2} \,,\  
\tens d^1_{1,1}(\beta) = \frac{1+\cos\beta}{2}
\label{eq.dinit}
\end{equation}
and the recurrence relation \citep{Gimbutas_JCP_2009}
\begin{equation}
\begin{split}
\tens d^l_{m,m'}(\beta) =& 
 +\sqrt{\frac{(l+m')(l+m'-1)}{(l+m)(l+m-1)}} \tens d^1_{1,1}(\beta) \tens d^{l-1}_{m-1,m'-1}(\beta)
\\ &
- \sqrt{\frac{(l+m')(l-m'  )}{(l+m)(l+m-1)}} \sin(\beta) \tens d^{l-1}_{m-1,m'}(\beta)
\\ &
+ \sqrt{\frac{(l-m')(l-m'-1)}{(l+m)(l+m-1)}} \tens d^1_{1,-1}(\beta) \tens d^{l-1}_{m-1,m'+1}(\beta)
\end{split}
\label{eq.drec}
\end{equation}
which also implies
\begin{equation}
\tens d^l_{l,l}(\beta) = \tens d^1_{1,1}(\beta) \tens d^{l-1}_{l-1,l-1}(\beta)
\qquad \text{and} \qquad
\tens d^l_{l,-l}(\beta) = \tens d^1_{1,-1}(\beta) \tens d^{l-1}_{l-1,1-l}(\beta)\ .
\label{eq.drecb}
\end{equation}
The algorithm is the following:
\begin{algorithm}[H]
\caption{\label{algo}Calculate Wigner d matrix}
\begin{algorithmic}
\STATE {initialize $\tens d^1_{0,0}$, $\tens d^1_{1,1}$, $\tens d^1_{1,0}$, and
$\tens d^1_{1,-1}$ from (Eq.~\ref{eq.dinit})}
\STATE {calculate the other terms of order $1$ using the symmetries (Eq.~\ref{eq.dsym})}
\FOR{$l=2$ to $l_\mathrm{max}$}
\STATE \mycomment{apply the recurrence relations as follows}
\FOR{$m=0$ to $l$}
\FOR{$m'=\MAX(-m,1-l)$ to $\MIN(m,l-1)$}
\STATE {calculate $\tens d^l_{m,m'}$ using (Eq.~\ref{eq.drec})}
\ENDFOR
\ENDFOR
\STATE {compute $\tens d^l_{l,l}$ and $\tens d^l_{l,-l}$ from (Eq.~\ref{eq.drecb})}
\STATE {calculate the other terms of order $l$ using the symmetries (Eq.~\ref{eq.dsym})}
\ENDFOR
\end{algorithmic}
\end{algorithm}

For completeness, we also provide the explicit terms at order $l=2$ in
Table~\ref{tab.wigner}.

\begin{table}[h!]
\begin{center}
\caption{\label{tab.wigner}Explicit Wigner d matrix $\tens d^2_{m,m'}(\beta)$}
\renewcommand{\arraystretch}{1.8}
\begin{tabular}{c|c|c|c|c|c}  \hline\hline
\tikz{\clip (-0.25,-0.21) rectangle (0.33,0.35);
      \path (0.2,0.2) node {$m'$};
      \path (-0.16,-0.16) node {$m$};
      \draw (-0.25,0.2) -- (0.3,-0.2);}
& 2 & 1 & 0 & -1 & -2 \\ \hline 
& & & & & \\[-1.5em]
2 
& \mytiny{$\displaystyle\frac{(1+\cos\beta)^2}{4}$                  }
& \mytiny{$\displaystyle-\frac{\sin\beta(1+\cos\beta)}{2}$          }
& \mytiny{$\displaystyle\frac{1}{2}\sqrt{\frac{3}{2}} \sin^2\beta$  }
& \mytiny{$\displaystyle-\frac{\sin\beta(1-\cos\beta)}{2}$          }
& \mytiny{$\displaystyle\frac{(1-\cos\beta)^2}{4}$                  } \\[0.5em]
1
& \mytiny{$\displaystyle\frac{\sin\beta(1+\cos\beta)}{2}$           }
& \mytiny{$\displaystyle\frac{2\cos^2\beta+\cos\beta-1}{2}$         }
& \mytiny{$\displaystyle-\sqrt{\frac{3}{2}}\sin\beta\cos\beta$      }
& \mytiny{$\displaystyle-\frac{2\cos^2\beta-\cos\beta-1}{2}$        }
& \mytiny{$\displaystyle-\frac{\sin\beta(1-\cos\beta)}{2}$          } \\[0.5em]
0
& \mytiny{$\displaystyle\frac{1}{2}\sqrt{\frac{3}{2}} \sin^2\beta$  }
& \mytiny{$\displaystyle\sqrt{\frac{3}{2}}\sin\beta\cos\beta$       }
& \mytiny{$\displaystyle\frac{3\cos^2\beta-1}{2}$                   }
& \mytiny{$\displaystyle-\sqrt{\frac{3}{2}}\sin\beta\cos\beta$      }
& \mytiny{$\displaystyle\frac{1}{2}\sqrt{\frac{3}{2}} \sin^2\beta$  } \\[0.5em]
-1
& \mytiny{$\displaystyle\frac{\sin\beta(1-\cos\beta)}{2}$           }
& \mytiny{$\displaystyle-\frac{2\cos^2\beta-\cos\beta-1}{2}$        }
& \mytiny{$\displaystyle\sqrt{\frac{3}{2}}\sin\beta\cos\beta$       }
& \mytiny{$\displaystyle\frac{2\cos^2\beta+\cos\beta-1}{2}$         }
& \mytiny{$\displaystyle-\frac{\sin\beta(1+\cos\beta)}{2}$          } \\[0.5em]
-2
& \mytiny{$\displaystyle\frac{(1-\cos\beta)^2}{4}$                  }
& \mytiny{$\displaystyle\frac{\sin\beta(1-\cos\beta)}{2}$           }
& \mytiny{$\displaystyle\frac{1}{2}\sqrt{\frac{3}{2}} \sin^2\beta$  }
& \mytiny{$\displaystyle\frac{\sin\beta(1+\cos\beta)}{2}$           }
& \mytiny{$\displaystyle\frac{(1+\cos\beta)^2}{4}$                  } \\[0.5em] \hline
\end{tabular}
\end{center}
\end{table}

\section{Time derivatives}
\label{sec.td}
Let a function $f(\vv x, t)$ developed in spherical harmonics as
\begin{equation}
f(\vv x, t) = \sum_{l,m} \bar z_{l,m}(t) Y_{l,m}(\hvec x)
\end{equation}
in the inertial frame ${\cal F}_0$, and as
\begin{equation}
f(\vv x, t) = \sum_{l,m} \bar Z_{l,m}(t) Y_{l,m}(\hvec X)
\end{equation}
in the body frame ${\cal F}_p$. For any constant vector $\vv x$ in ${\cal F}_p$, we have
\begin{equation}
\dot{\vec x} = \vec \omega \times \vec x
\qquad \mathrm{and} \qquad
\dot{\vec X} = \vec 0 \ ,
\end{equation}
with respect to the frame ${\cal F}_p$.
By consequence, in ${\cal F}_p$, on the one hand,
\begin{equation}
\dot f(\vv x, t) = \sum_{l,m}
\left(
\dot {\bar z}_{l,m}(t)
Y_{l,m}(\hvec x)
+ \bar z_{l,m}(t) \dot{\vec x}\cdot \vec \nabla Y_{l,m}(\hvec x)
\right)
\ ,
\label{eq.dfx}
\end{equation}
and on the other hand,
\begin{equation}
\dot f(\vv x, t) = \sum_{l,m}
\dot {\bar Z}_{l,m}(t) Y_{l,m}(\hvec X)\ .
\label{eq.dfX}
\end{equation}
But given that the time derivative of $\vec x$ is $\dot{\vec x} = \vec\omega\times\vec x$, we get 
\begin{equation}
\dot{\vec x}\cdot\vec \nabla =  
(\vec \omega \times \vec x) \cdot \vec \nabla = 
\ii (\vec \omega \cdot \vec J)
\label{eq.dXnabla}
\end{equation}
where $\vec J = -\ii\vec x \times \vec \nabla$ is the angular momentum
operator and where, by construction of the scalar product
\citep{Varshalovich_book_1988},
\begin{equation}
\vec \omega \cdot \vec J = - \omega_{+} J_{-} + \omega_0 J_0 - \omega_{-} J_{+}\ .
\end{equation}
We then define a matrix $\tens J(\vec \omega)$ of size $(2l+1)\times(2l+1)$ such that
\begin{equation}
(\vec \omega \cdot \vec J) Y_{l,m}(\hvec x) = \sum_{m'=-l}^l [\tens J(\vec
\omega)]^l_{m',m} Y_{l,m'}(\hvec x)\ ,
\label{eq.matJ}
\end{equation}
where all non-zero coefficients are
\begin{equation}
\begin{array}{lll}
\left[\tens J(\vec \omega)\right]^l_{m-1,m} &=& - \sqrt{\frac{l(l+1)-m(m-1)}{2}} \omega_+\ , \\
\left[\tens J(\vec \omega)\right]^l_{m,m}   &=& m\, \omega_0 \ , \\
\left[\tens J(\vec \omega)\right]^l_{m+1,m} 
&=& + \sqrt{\frac{l(l+1)-m(m+1)}{2}} \omega_-\ .
\end{array}
\end{equation}
Combining Eqs. (\ref{eq.Wigner}), (\ref{eq.dfx}-\ref{eq.dXnabla}), and
(\ref{eq.matJ}), we obtain
\begin{equation}
\sum_{m'} \tens D^l_{m,m'}\dot {\bar Z}_{l,m'} = 
\dot {\bar z}_{l,m} + \ii \sum_{m'}\,[\tens{J}(\vec \omega)]^l_{m,m'} {\bar z}_{l,m'}
\ .
\end{equation}

\section{Fourier transform}
\label{sec.Fourier}
Let two functions $f(\vv x, t)$ and $g(\vv x, t)$ expanded in spherical
harmonics as $f = \sum_l f_l$ and $g=\sum_l g_l$ with
$$
f_l(\vv x, t) = \sum_{m=-l}^l \bar Z_{l,m}(t) Y_{l,m}(\hvec X)
\qquad \text{and} \qquad
g_l(\vv x, t) = \sum_{m=-l}^l \bar Z'_{l,m}(t) Y_{l,m}(\hvec X)
$$
in the frame ${\cal F}_p$ and
$$
f_l(\vv x, t) = \sum_{m=-l}^l \bar z_{l,m}(t) Y_{l,m}(\hvec x)
\qquad \text{and} \qquad
g_l(\vv x, t) = \sum_{m=-l}^l \bar z'_{l,m}(t) Y_{l,m}(\hvec x)
$$
in ${\cal F}_0$. Let $\alpha$, $\beta$, and $\gamma=\omega t$ be the three
angles such that
$$
\vec x = \tens R_3(\alpha) \tens R_2(\beta) \tens R_3(\gamma) \vec X\ .
$$
We have then
\begin{equation}
z_{l,m}(t) = \sum_{m'=-l}^l \bar{\tens D}^l_{m,m'}(t) Z_{l,m'}(t)
\qquad \text{and} \qquad
z'_{l,m}(t) = \sum_{m'=-l}^l \bar{\tens D}^l_{m,m'}(t) Z'_{l,m'}(t)
\label{eq.zZ}
\end{equation}
with
$$
\tens D^l_{m,m'} (t) = \tens D^l_{m,m'}(0) \e^{-\ii m' \omega t}\ .
$$
Let us further assume that the two functions are
related to each other in ${\cal F}_p$ by
$$
f_l(\vv x, t) = h_l(t) * g_l(\vv x, t)
\qquad \text{for all $l$}\ ,
$$
where $h_l(t) \in \mathbb{R}$ is a real distribution. The symbol $*$
denotes the convolution product. As the convolution is done with respect
to time, the orthogonality of the spherical harmonics implies that for
all $l$ and $m$,
\begin{equation}
Z_{l,m}(t) = h_l(t) * Z'_{l,m}(t)\ .
\label{eq.Zconv}
\end{equation}
Combining Eqs.~(\ref{eq.zZ}) and (\ref{eq.Zconv}), we get
\begin{align}
z_{l,m}(t) &= \sum_{m'=-l}^l \sum_{m"=-l}^l \int_{-\infty}^
\infty \bar{\tens D}^l_{m,m'}(t) h_l(t-t') {\tens D}^l_{m",m'}(t')
z'_{l,m"}(t')\,dt' \nonumber \\
&= \sum_{m"=-l}^l \tens h^l_{m,m"}(t) * z'_{l,m"}(t)\ ,
\label{eq.zconvtens}
\end{align}
where
$$
\tens h^l_{m,m"}(t) = \sum_{m'=-l}^l \bar{\tens D}^l_{m,m'}(0) h_l(t) \e^{\ii m'
\omega t} \tens D^l_{m",m'}(0)\ .
$$
In particular, if the rotation axis $\vv \omega$ is aligned with the
third axis of ${\cal F}_0$ and ${\cal F}_p$, i.e., if $\alpha=\beta=0$,
$\tens h^l_{m,m"}(t)$ is diagonal and we obtain
\begin{equation}
z_{l,m}(t) = \left(h_l(t)\e^{\ii m \omega t}\right) * z'_{l,m}(t)
\qquad\text{if\quad$\alpha=\beta=0$}\ .
\label{eq.zconvdiag}
\end{equation}
Taking the Fourier transform of Eqs.~(\ref{eq.zconvtens}) and
(\ref{eq.zconvdiag}), we get
$$
\ubar z_{l,m}(\nu) = \sum_{m"=-l}^l \ubar {\tens h}^l_{m,m"}(\nu) \ubar z'_{l,m"}(\nu)
\quad
\text{with}
\quad
\ubar{\tens h}^l_{m,m"}(\nu) = \sum_{m'=-l}^l \bar{\tens D}^l_{m,m'}(0)
\ubar h_l(\nu-m'\omega) \tens D^l_{m",m'}(0)\ ,
$$
on the one hand, and
$$
\ubar z_{l,m}(\nu) = \ubar h_l(\nu - m\omega) \ubar z'_{l,m}(\nu)
\qquad\text{if\quad$\alpha=\beta=0$}\ ,
$$
on the other.

\bibliographystyle{spbasic}      % basic style, author-year citations
\bibliography{maxwell3d}   % name your BibTeX data base

\end{document}